\documentclass[12pt]{article}
\usepackage{epsf}
\usepackage{exscale}
\usepackage{amsmath}
\usepackage{graphicx}

\def\la{\mathrel{\mathpalette\fun <}}

\def\fun#1#2{\lower3.6pt\vbox{\baselineskip0pt\lineskip.9pt
\ialign{$\mathsurround=0pt#1\hfil##\hfil$\crcr#2\crcr\sim\crcr}}}

\begin{document}

\titlepage
\title{\begin{flushright}
Preprint PNPI-2795,  2009
\end{flushright}
\bigskip\bigskip\bigskip
Resonance capture by hydrogenous impurities and losses of
ultracold neutrons in solid material traps}
\date{}
\author{G.S.
Danilov \thanks{E-mail address:  danilov@thd.pnpi.spb.ru}\\ Petersburg
Nuclear Physics Institute,\\ Gatchina, 188300, St.-Petersburg, Russia
\\}
\maketitle
\begin{abstract}
The capture of trapped ultracold neutrons (UCNs) by closed hydrogenous
impurities within a solid coating of  the trap is discussed as
a possible cause of observed anomalously large losses of UCNs in solid
material UCN traps. Then significant losses of UCNs arise only
if resonances occur in the UCN-impurity scattering amplitude.
For a large size impurity,
higher partial waves in the UCN-impurity interaction are important, and
they are taken into account in the present paper.  The method of
the calculation is applicable to irregular shape impurities as well. A
small distortion of an impurity shape, if it splits the resonance, can
increase the UCN losses by a few times.  UCN losses in the
beryllium trap are calculated
assuming they are due to the UCN capture by ice
spherical impurities within the coating of the trap walls.
Both s- and p-wave
resonances contribute significantly to the UCN losses considered.
As an example, observed anomalous large UCN losses are achieved if
the average radius of the impurity is about 600 \AA$\,$ and the impurity
density is about $3\times 10^{14}$/cm$^3$.  A
distortion of the spherical shape of the impurity could
increase the UCN losses and therefore decrease the impurity density.

\end{abstract}
\newpage

\section{Introduction}

Losses of ultracold neutrons (UCNs) in material traps have attracted
attention in recent years \cite{serebrov}. The matter arises for crystal
materials such as
beryllium and graphite \cite{serebrov,serebrov92,arzum} especially for
beryllium \cite{serebrov,serebrov92}.  For a liquid or  solid  Fomblin
oil the observed losses could be explained by inelastic scattering
processes \cite{inel}, but it can not \cite{serebrov,arzum,inel,inelt}
explain the UCN losses in beryllium traps.  Indeed, the UCN losses in
beryllium traps depend slowly on the temperature $T$ at low temperature
\cite{serebrov,serebrov92,arzum}.  By contrast,
the inelastic scattering
should lead to a $\sim T^3$ dependence of the UCN losses.  The
discussed losses cannot be a result of  a surface heating caused by
hydrogenous surface impurities \cite{serebrov} because  it requires
too high a concentration of the hydrogen on the surface. In
addition, in this case the losses should be significantly reduced by
a degassing procedure, but  the observed UCN ones are insensitive to it
\cite{serebrov,serebrov92,arzum}.
The discussed UCN losses cannot be from a
coherent absorption of UCN by beryllium because the UCN capture cross
section for beryllium is too small.  The scattering of UCNs
by vacuum cavities can increase the UCN losses \cite{ignat}, but it
can hardly  explain the  anomalously large losses observed
\cite{serebrov,serebrov92}.

In the present paper we consider UCN losses from the UCN  capture
by closed impurities within the
solid coating of the trap walls. A low temperature  of the trap
is assumed so that only the elastic scattering and
capture of UCNs can take place. Ice impurities are mainly kept in mind
because the neutron-hydrogen capture cross section is large.
Water within a closed cavity can hardly be removed by a
degassing procedure.  So losses are insensitive to the degassing
process and do not depend on the trap
temperature, nearly as it has been
observed experimentally.

A possible correlation of large losses of UCNs in
beryllium with incoherent processes
has been noted in \cite{serebrov,serebrov1}. In \cite{serebrov}
the concept of a localization
of UCNs around the lattice defect has been proposed
which drasticly increases the UCN losses.
This mechanism of  UCN
losses  is not, however, acceptable since it
is based on a
mistakable solution of the scattering problem \cite{barab}.

At low
temperatures the losses of UCNs arise solely when absorption of UCNs
takes place.
If UCN losses are due to
coherent absorption of UCN by the medium,  then
the well known expression \cite{ignat} for
the coefficient $\mu(y)$ of the UCN losses is as  follows:
\begin{equation}
\mu(y)= \frac{2\eta}{y^2}(\arcsin y-y\sqrt{1-y^2}),\quad
y=\sqrt{E/U'}\,,
\label{intro}
\end{equation}
where $E$ is the UCN
energy, $U'$ is the real part of the optical potential, and $\eta=
U''/U'$, where $U''$ is the imaginary part of the optical potential of
the medium. Inelastic scattering is negligible at low temperatures
and so $\eta=b^\prime/b$, where $b^\prime$ is the UCN absorption amplitude and $b$
is the UCN elastic scattering amplitude.  Then for beryllium
$\eta=2.47\times10^{-7}$ while the experimental losses \cite{serebrov}
require $\eta=3\times10^{-5}$.  This exceeds the expected value of
$\eta$ by a factor of about 100.  The UCN losses due to  UCN capture
by a small size impurity are again represented by (\ref{intro}) with
the understanding  that $\eta$ is replaced by $p_V w$ where $p_V$ is
the portion of the total volume occupied by impurities, and $w=W/U'$.
Here $(-W)$ is the imaginary part of the UCN- impurity potential.  For
hydrogen $w$ is known to be $w\approx 3\times10^{-6}$.  Then
$\eta\sim10^{-5}$ requires  $p_V\sim10$ which is an obviously
unacceptable result.

The losses
drastically grow for  a large size impurity when a
resonance in the UCN-impurity scattering amplitude occurs.
Compared with the UCN losses
by a vacuum cavity \cite{ignat}, they are
increased by a factor, which roughly is
$w a\kappa U'/(E\eta)$. In this case $a$ is the radius of
the impurity and $\kappa=\sqrt{2m(U-E)/\hbar^2}$  is the length of the
UCN penetration into the coating; $m$ is the neutron mass and $\hbar$
is the Plank constant (reduced). Under the discussed conditions this
factor is $\sim 100$.

As is shown in the paper, a relatively small impurity shape
distortion, which splits the resonance into $N$ resonances,
increases the UCN losses, roughly, by N times.  In particular, a small
distortion of the spherical impurity is able to increases the UCN
losses from the  $l$-wave resonance by about $(2l+1)$ times in
comparison with the UCN losses from a spherical impurity occupying the
same volume.  In the paper, however, as the first step, spherical shape
impurities are mainly considered.

Computations are performed under conditions
where experimental data of the UCN losses
are obtained \cite{serebrov}. Then the maximal UCN energy in
the trap is about 23 cm, 38 cm, 46 cm, 52 cm and 58 cm (this means that
the UCN energy is the same as the gravitational energy of the neutron
lifted to the given height; 1 cm energy corresponds to
$1.025\times10^{-9}$ eV).  The radius of ice impurity is taken in the
range (453.7 -- 875.2) \AA.  If the radius is less than 464.2
\AA, then a resonance does not occur for the UCN energy below 58 cm,
and such size impurities do not contribute to the UCN losses. In the
radius range being approximately (460--740) \AA, only an s-wave
resonance occurs, and solely the s-wave UCN-impurity interaction is
important.  If the radius is larger than 740 \AA, then both s- and
p-wave interactions contribute to the UCN losses. The $s-p$ wave
interference is taken into account, too.  The UCN losses mainly arise
from impurities lying rather far from the trap wall, but not from those
lying on the trap surface.  The d-wave interaction is negligible
since a $d$-wave resonance does not occur in the discussed
energy-radius region.

When an $l$-wave resonance is present, then the $l$-wave nonresonance
background grows and contributes about (20--30)$\%$ to the UCN losses.
Therefore a resonance approximation such as that employed in
\cite{ignat} is not used in the above computations. A large radius
approximation \cite{ignat} is not used in above computations, too.
Nevertheless, both these approximations are discussed  (in Sec. IV)
for a semiquantitative consideration of the matter.

An  example of the
distribution of the ice impurities is proposed fitting the
experimental data \cite{serebrov}.
Now experimental data are insufficient to fit them in the unique way.
It assumes that
no  impurities have  radii greater than 875 \AA.
The impurity density
is required to be about $3\times 10^{14}$/cm$^3$. In fact the
density falls when the size of the impurity increases. As a
result, the average size of the impurity is about 600 \AA.
The impurities occupy 0.25 of the total volume. This is
rather a large part of the volume, but
not so much that
the UCN capture by hydrogenous impurities should  be discarded as a
possible cause of the UCN losses.  Perhaps, in accordance with
aforesaid, distortions of the spherical impurity  are able to
decrease the volume occupied by impurities.
At low temperatures the hydrogenous capture of UCNs
remains to be a solely  plausible cause for UCN losses, and so
a knowledge of  the cavity distribution in shape,
orientation and size would be important. The
method of this paper allows to perform calculations for
more  realistic  impurity distributions.

Under resonance conditions, the
UCN-impurity scattering amplitude is able to achieve  extremely
large magnitudes $\sim 100$km! In spite of this,
the  UCN can be treated as interacting with a single isolated
impurity.  Indeed,
a macroscopic effect from an interference of scattered waves
each being formed by its own scatterer, could arise if  many
scattered waves interfered with the given scattered wave. Hence the
effect  could be mainly  due to impurities separated from
each other by a macroscopic scale distance $\sim\widetilde L$ in directions
parallel to the boundary of the trap. Simultaneously, a sub-barrier
scattered wave exponentially decreases going away from the scatterer. As
the result, the discussed interference
effect is negligible.  Another effect could be that the scattered
wave being formed by a given impurity, falls on another impurity
after it was reflected from the trap boundary.
The possibility  of such
a process decreases as $1/\widetilde L^2$ independently of the magnitude of
the scattering amplitude. So this process is negligible, too.  This is
demonstrated in Appendix A of the paper.

The paper is organized as it follows. In Sec. II the wave function of
the UCN in the trap is calculated. Inside the material of the trap it
is represented as a superposition of wave functions $\psi_0$.  Each
$\psi_0$ is  the analytical continuation to  sub-barrier UCN energies
of the wave function of a neutron in the infinite homogeneous matter
with impurities.  By this construction, the wave function of the UCN in
the trap is already matched on the boundary of the impurity. Matching
on the trap boundary leads to a linear integral equation for the UCN
wave function in the trap through $\psi_0$. The result is obtained
without any restrictions on the UCN-impurity interaction. This is used
in Appendix A to estimate interference effects discussed in the
previous paragraph.  They being negligible, $\psi_0$ is given in terms
of the relevant amplitude of the UCN scattering by a single isolated
impurity in the infinite homogeneous matter, and so the UCN losses
are calculated through this scattering amplitude, too. It is similar
to the way in which the incoherent losses due to $s$-wave scattering of
the UCN by the vacuum cavity have been calculated \cite{ignat} in terms
of the $s$-wave amplitude of the UCN-vacuum cavity scattering.

In Sec. III the results of Sec. II
are discussed for the spherical shape
impurity case.

In Sec. IV   interactions of the UCN with a very small and a large
size impurity are considered. In the large size impurity approximation,
a semiquantitative consideration of the UCN losses is given, and a
comparison with the case of the UCN scattering by a vacuum cavity
\cite{ignat}, is performed.

In Sec. V a small quadruple
distortion shape impurity is discussed.

In Sec. VI the UCN numerical computations are given,
some details being given in
Appendix B.

\section{Interaction of UCN with impurities}

In this Section
a wave function of UCN in the material trap
is calculated. Expressions for UCN loss cross sections and for
coefficients of the UCN losses are given.  It is implied that UCN
interacts with material impurities, which presumably  exist inside the
coating of the trap walls.  Basic results of this Section are
applicable without restrictions on the UCN-impurity interaction. A set
of impurity parameters is denoted as $\{\alpha\}$.  If UCN interacts
 with a spherical single impurity, then $\{\alpha\}=({\bf r_c},a)$,
where ${\bf r_c}$ is a radius-vector of the impurity center and $a$
is the impurity radius.  It is accepted that the plane $z=0$ is the
border between the vacuum space ($z<0$) and a material medium  ($z>0)$
containing impurities.

The UCN wave
function $\psi_1({\bf r},{\bf k_0}, \{\alpha\})$
in the vacuum  space is given by
\begin{equation}
\psi_1({\bf r},{\bf k_0},
\{\alpha\})
=e^{i{\bf k_0}\cdot
{\bf r}}+R(q_0)
e^{-i\widehat k_0(q_0)z+i\overrightarrow{ q_0}\cdot\overrightarrow{ l}}
+\int C_1
(\overrightarrow{ q},{\bf k_0},
\{\alpha\})
e^{i\overrightarrow{ q}\cdot
\overrightarrow{ l}}
e^{-i\widehat k_0(q)z}d^2q\,,
\label{ps1il}
\end{equation}
where ${\bf r}=
(\overrightarrow{ l},z)=(x,y,z)$ is the radius-vector of the space point in
question. In this case $\overrightarrow{l}$ is a two-dimensional (2D) vector  in
$XY$-plane.  Further, ${\bf k_0}$ is the UCN wave vector:  ${\bf
k_0}=(\overrightarrow{ q_0},\, \widehat k_0(q_0))$ where $\overrightarrow{ q_0}$ is a 2D
vector in $XY$-plane.  So $q_0=k_0\sin\theta$, where $\theta$ is a hide
angle. Furthermore, $R(q_0)$ is the well known coefficient of the
coherent reflection of UCN from the border,
\begin{equation}
R(q)=\frac{\widehat k_0(q)-i\widehat\kappa(q)}{\widehat
k_0(q)+i\widehat\kappa(q)}\,.
\label{reflco}
\end{equation}
In (\ref{ps1il}) and (\ref{reflco}) and throughout the
paper,
\begin{equation}
\widehat k_0^2(q)=k_0^2-q^2\,,\quad
k_0^2=2mE/\hbar^2\,,\quad
\kappa_0^2=2mU/\hbar^2\,,\quad
\widehat\kappa^2(q)=\kappa^2+q^2,\quad
\kappa^2=\kappa_0^2-k_0^2\,,
\label{hatkkl}
\end{equation}
where $m$ is the
neutron mass, and
$U$ is the optic potential of the material medium;
$ReU>0$ and $ImU\leq0$.
The sub-barrier case $E<ReU$ is considered.
The first term in the right side of (\ref{ps1il})
describes the falling wave, the second term represents
the wave formed due to
the coherent reflection of UCN from the border, and the integral
describes the incoherent wave due to the UCN-impurity
interaction.
The integration  is performed  from $q=0$  to $q\to\infty$.
The condition $q>k_0$ determines the intrinsic
reflection region. In this case $\widehat k_0(q)=i|\widehat k_0(q)|$.

The wave function  (\ref{ps1il}) is matched
at $z=0$ with the wave
function $\psi_2({\bf r},{\bf k_0},
\{\alpha\})$
in the material medium  off impurities as  follows:
\begin{equation}
\psi_1({\bf r},{\bf k_0},
\{\alpha\})=\psi_2({\bf r},{\bf k_0},
\{\alpha\})\biggl|_{z=0},\quad
\partial_z\psi_1
({\bf r},{\bf k_0},
\{\alpha\})=\partial_z\psi_2(
{\bf r},{\bf k_0},
\{\alpha\})\biggl|_{z=0}\,.
\label{sew}
\end{equation}
A full set of boundary conditions includes,
in addition, the matching of  wave functions on the impurity boundary.
An important subtlety of the task is that
$\psi_2({\bf r},{\bf k_0},
\{\alpha\})$ is
represented by relatively a simple superposition
of functions
$\psi_0({\bf r},\overrightarrow{ p},\kappa,
\{\alpha\})$  each is
an analytical
continuation from $(E-U)>0$ to $(E-U)<0$ of
the wave function of the neutron
in the material medium with
impurities:
\begin{equation}
\psi_0({\bf r},\overrightarrow{ p},\kappa,
\{\alpha\})= e^{i{\bf b}\cdot
{\bf r}}+\widetilde
\psi_0({\bf r},\overrightarrow{ p},\kappa,
\{\alpha\})\,,\quad
{\bf b}=(\overrightarrow{ p}\,,i\widehat\kappa(p))\,,\quad b^2=-\kappa^2\,,
\label{infmat}
\end{equation}
where the first term in the right side describes the falling wave,
and the second term arises due to the scattering of the neutron by
impurities.
In this case $\overrightarrow{ p}=(p_x,p_y)$ is the 2D vector in
$XY$-plane, $p_x$ and $p_y$ being components of the UCN momentum, and
$\kappa$ is defined in (\ref{hatkkl}).
Since the
reflection from the border does not change the UCN energy, all
$\psi_0({\bf r},\overrightarrow{ p},\kappa,
\{\alpha\})$ correspond to the same
neutron energy $E=U-\hbar\kappa^2/2m$, but they are distinguished in
$\overrightarrow{p}$. Therefore
\begin{equation}
\psi_2({\bf r},{\bf k_0},
\{\alpha\})=\int C_2
(\overrightarrow{ p},{\bf k_0},\{\alpha\})
\psi_0({\bf r},\overrightarrow{ p},\kappa,
\{\alpha\})
d^2p\,.
\label{matwf}
\end{equation}
If UCN interacts with a spherical, single
impurity, then  $\widetilde
\psi_0({\bf r},\overrightarrow{ p},\kappa,
\{\alpha\})$ in (\ref{infmat}) is as follows:
\begin{equation}
\widetilde\psi_0({\bf r},\overrightarrow{ p},\kappa,
\{\alpha\})= e^{i{\bf b}\cdot
{\bf r_c}}F(\kappa,a,-i{\bf b}\cdot
{\bf \nabla}/b^2)\frac{e^{ib|{\bf r}-
{\bf r_c}|}}{|{\bf r}-
{\bf r_c}|}\,,
\label{tfmat}
\end{equation}
where, as before, ${\bf r_c}$ is
the radius-vector of the impurity center and $a$ is
the impurity radius.
Furthermore,
$F(\kappa,a,\cos\theta)$ is the scattering amplitude, $\theta$
being scattering angle,
$\cos\theta={\bf b}\cdot{\bf b}^\prime/b^2$. In this case
${\bf b}^\prime$ is the final state wave vector
($b=b^\prime=i\kappa$) which was  replaced in (\ref{infmat}) by its operator
$-i{\bf \nabla}$. Outside the impurity  the wave
function (\ref{infmat}) satisfies to the Schr\"{o}dinger equation
with the U interaction potential and
has the true asymptotics at $r\to\infty$ as follows:
\begin{equation}
\psi_0({\bf r},\overrightarrow{ p}, \kappa, \{\alpha\})\to
e^{i{\bf b}\cdot{\bf r}} + e^{i{\bf b}\cdot
{\bf r_c}}F(\kappa,a,
{\bf b}{\bf b}^\prime/b^2)
\frac{e^{ib|{\bf r}-
{\bf r_c}|}}{|{\bf r}-
{\bf r_c}|}\,.
\label{afmat}
\end{equation}
The factor $\exp(i{\bf b} {\bf r_c})$ in
the second term in the right side of
(\ref{infmat}) is the value of the
plane wave in the center of the impurity.  As it usually is, the
scattering amplitude is expanded over partial waves amplitudes which
are well known for a constant interaction potential \cite{landau}.

Eq.(\ref{tfmat}) is
directly extended to the case of an arbitrary shape impurity. Then
${\bf r_0}$  determines a certain convenient point inside
the impurity and $F(\kappa,a,\cos\theta)$ is replaced by a
relevant scattering amplitude $F({\bf
b\,,b^\prime},\{\alpha\})$.

If $\psi_0({\bf r},\overrightarrow{ p},\kappa,
\{\alpha\})$
is known,
or if a reliable approximation to this can be proposed, then $C_1 (\overrightarrow{
q},{\bf k_0}, \{\alpha\})$    and $C_2 (\overrightarrow{ p},{\bf k_0},\{\alpha\})$
are  calculated from boundary conditions (\ref{sew}).
For this purpose
$\widetilde
\psi_0({\bf r},\overrightarrow{ p},\kappa,
\{\alpha\})$ in (\ref{matwf}) is represented
by means of
Fourier integral over  two-dimensional vector $\overrightarrow{ q}$
as  follows:
\begin{equation}
\widetilde
\psi_0({\bf r},\overrightarrow{ p},\kappa,
\{\alpha\})=\int e^{i\overrightarrow{ q}\cdot
\overrightarrow{ l}} \widehat
\psi_0(\overrightarrow{ q},z,\overrightarrow{ p},\kappa,
\{\alpha\})\,d^2q\,,
\label{four}
\end{equation}
where definitions are given in (\ref{hatkkl}).
Inasmuch as (\ref{four})
obeys the Schr\"{o}dinger equation, then
$\widehat
\psi_0(\overrightarrow{ q},z,\kappa,\overrightarrow{ p},
\{\alpha\})$ depends on $z$ as
$\exp{(\pm \widehat\kappa(q) z)}$.
Since $\widetilde
\psi_0({\bf r},\overrightarrow{ p},\kappa,
\{\alpha\})$ is  a scattered wave formed
in the right half space ($z>0$), only $\exp{(\widehat\kappa(q)
z)}$ survives at $z\to0$.  Hence
\begin{equation}
\partial_z\widehat\psi_0(\overrightarrow{ q},z,
\overrightarrow{ p},\kappa,
\{\alpha\})=\widehat\kappa(q)\widehat \psi(\overrightarrow{ q},
\overrightarrow{ p},\kappa,
\{\alpha\})\biggl|_{z=0}\,,
\label{derz}
\end{equation}
where $\widehat
\psi(\overrightarrow{ q},\overrightarrow{ p},\kappa,
\{\alpha\})\equiv\widehat
\psi_0(\overrightarrow{ q},0,\overrightarrow{ p},\kappa,
\{\alpha\})$.
Then (\ref{sew}) is turned out to be
\begin{eqnarray}
[1+R(q_0)]\delta^2(\overrightarrow{ q}-\overrightarrow{ q_0})+
C_1(\overrightarrow{ q},{\bf k_0},\{\alpha\})
=\ C_2(\overrightarrow{ p},{\bf k_0},\{\alpha\})
+\widetilde B(\overrightarrow{ q}, {\bf k_0}, \{\alpha\})\,,
\nonumber\\
i\widehat k_0(q_0)[1-R(q_0)] \delta^2(\overrightarrow{ q}-\overrightarrow{ q_0})
-i\widehat k_0(q) C_1(\overrightarrow{ q},{\bf k_0},\{\alpha\})
\nonumber\\
 =-\widehat\kappa(q) C_2
(\overrightarrow{ p},{\bf k_0},\{\alpha\})
+\widehat\kappa(q)\widetilde B(\overrightarrow{ q},
{\bf k_0},\{\alpha\}) \,,
\label{sew1}
\end{eqnarray}
where
\begin{equation}
\widetilde B(\overrightarrow{ q},
{\bf k_0},\{\alpha\})=\int C_2
(\overrightarrow{ p},{\bf k_0},\{\alpha\})
\widehat\psi (\overrightarrow{ q},\kappa,\overrightarrow{ p},
\{\alpha\})\,d^2p\,,
\label{tif}
\end{equation}
and other definitions are given in (\ref{hatkkl}), (\ref{four})
and (\ref{derz}).  From (\ref{sew1}), it follows that
\begin{eqnarray}
C_2(\overrightarrow{ p},{\bf k_0},\{\alpha\})=
\frac{2\widehat k_0(q_0)}{\widehat k_0(q_0)+i\widehat\kappa(q_0)}
\delta^2(\overrightarrow{ q}-\overrightarrow{ q_0})
-
R(q)
\widetilde B(\overrightarrow{ q},
{\bf k_0},\{\alpha\})\,,
\nonumber\\
C_1(\overrightarrow{ q},{\bf k_0},
\{\alpha\})=
\frac{2i\widehat\kappa(q)}{\widehat k_0(q)+i\widehat\kappa(q)}
\widetilde B(\overrightarrow{ q},
{\bf k_0},\{\alpha\})\,,
\label{solsew}
\end{eqnarray}
where $R(q)$ is given by (\ref{reflco}).
The $\sim\delta^2(\overrightarrow{ q}-\overrightarrow{ q_0})$ term
in $C_2(\overrightarrow{ p},{\bf k_0},\{\alpha\})$ describes
the  wave penetrating into the matter from the
vacuum, and the rest term is due to the reflection
of the scattered wave from the
boundary.  Equation for $\widetilde
B(\overrightarrow{ q}, {\bf k_0},\{\alpha\})$
is derived by substituting (\ref{solsew}) to
(\ref{tif}) as follows:
\begin{equation}
\widetilde
B(\overrightarrow{ q}, {\bf k_0},\{\alpha\})= [1+R(q_0)]
\widehat\psi(\overrightarrow{ q},\overrightarrow{ q_0},\kappa,
\{\alpha\})
-\,
\int R(p)
\widetilde B(\overrightarrow{ p},{\bf k_0},\{\alpha\})
\widehat\psi(\overrightarrow{ q},\overrightarrow{ p},\kappa,
\{\alpha\})\,d^2p \,,
\label{tif1}
\end{equation}
where definitions are given in (\ref{reflco}) and
(\ref{derz}).

Once eq.(\ref{tif1}) has been solved, the UCN wave function is
calculated using (\ref{solsew}).  The total number $N_l({\bf k_0})$ of
neutrons lost per second is given by the integral of the UCN current
density over the $z=0$ plane. Indeed, by first principals
\cite{landau}, the change per second of the number of particles in  a
volume is given by the particle current going through the volume
boundary. In terms of the wave function (\ref{ps1il}) the current
density $j(\overrightarrow{ l}, {\bf k_0}, \{\alpha\})$ is given by
\begin{equation}
j(\overrightarrow{ l}, {\bf k_0},
\{\alpha\})= \frac{\hbar}{m}Im\biggl(\psi_1^*( {\bf r},{\bf k_0},
\{\alpha\})\partial_z\psi_1({\bf r},{\bf k_0},
\{\alpha\})\biggl)
\,,
\label{current}
\end{equation}
where  the right-top star denotes complex conjugation
and the right part of (\ref{current}) is calculated at $z=0$.
Therefore,
\begin{equation}
N_l({\bf
k_0},\{\alpha\})
=\int j(\overrightarrow{ l}, {\bf k_0},\{\alpha\})\,d^2l
=\hbar\widehat k_0(q_0)[(1-|R( q_0)|^2)]S/m
+N_{il}({\bf
k_0},\{\alpha\})\,,
\label{tnlost}
\end{equation}
where $S$ is the area
of the boundary, $R(q_0)$ is given by (\ref{reflco}), and other
definitions are given in (\ref{hatkkl}). The $\sim S$ term in the right
side of (\ref{tnlost}) gives the number of UCNs lost per second due to
coherent
absorption of UCNs by the medium. The term $N_{il}(
{\bf k_0},\{\alpha\})$ gives the number of UCNs lost per second
due to the UCN capture by impurities. Furthermore,
$N_{il}({\bf k_0},\{\alpha\})m/[\hbar\widehat
k_0(q_0)]$ is the cross section $\sigma_c({\bf k_0},
\{\alpha\})$ of incoherent losses of UCNs in the trap.
Indeed, by  definition \cite{landau}, the cross section of the process
is the number of the questioned events per second being divided by
the density of the falling flow. Using
(\ref{ps1il}) and (\ref{current}), one obtains
$\sigma_c({\bf k_0},
\{\alpha\})$
as  follows
\begin{equation}
\sigma_c({\bf k_0},
\{\alpha\})=8\pi^2Im[R^*(q_0)(-i)
C_1
(\overrightarrow{ q},{\bf k_0},
\{\alpha\})]
- 4\pi^2
\int_{q<k_0} \frac{\widehat k_0(q)}{\widehat k_0(q_0)}|C_1
(\overrightarrow{ q},{\bf k_0},
\{\alpha\})|^2
d^2q\,.
\label{capt}
\end{equation}
The last term in the right side of (\ref{capt})
represents
the cross section $\sigma_s({\bf k_0}
\{\alpha\})$ of
the UCN incoherent scattering to the trap. Thus the first term
is the total cross section of incoherent processes \footnote{It is a
certain mishmash in \cite{ignat} where  the cross section in
the left side of
(\ref{capt}) is treated as the total cross section of
incoherent processes,
see eqs.  (8) and (9) of appendix 6.13 in
\cite{ignat}.}. So (\ref{capt}) is an optical theorem for incoherent
processes under the discussed conditions.  Eq.(\ref{capt}) can be also
written down using an amplitude $A_{s}(\overrightarrow{ q},{\bf k_0}, \{\alpha\})$
of the UCN incoherent scattering defined as follows:
\begin{equation}
A_{s}(\overrightarrow{ q},{\bf k_0}, \{\alpha\}) = -2i\pi\widehat k_0(q) C_1 (\overrightarrow{
q},{\bf k_0}, \{\alpha\})\,.
\label{scata}
\end{equation}
Taking into
account that
\begin{equation}
q=k_0\sin\theta,\qquad
d^2q=k_0^2cos\theta\,\sin\theta\, d\theta d\phi\,,
\label{qangd}
\end{equation}
$\sigma_s({\bf k_0}, \{\alpha\})$ can be represented as
follows:
\begin{equation}
\sigma_s({\bf k_0}, \{\alpha\})=\int
|A_{s}(\overrightarrow{ q},{\bf k_0}, \{\alpha\})|^2\,\frac{k_0}{\widehat
k_0(q_0)}\,d\Omega_s\,,
\label{crsecs}
\end{equation}
where integration
is performed over the spatial  angle $\Omega_s$ of the
scattering of UCN into the trap ($0<\theta<\pi/2$).  The factor
$k_0/\widehat k_0(q_0)$ arises in (\ref{crsecs}) because the velocity of
the passing of UCN through the boundary is $\hbar\widehat
k_0(q_0)/m$ while the velocity of the returned neutron is $\hbar
k_0/m$.  Then (\ref{capt}) is represented as follows:
\begin{equation}
\sigma_c({\bf k_0},
\{\alpha\})= \frac{4\pi}{\widehat k_0(q_0)}
Im[R^*(q_0)A_{s}({\bf k_0},
\{\alpha\})]
-\sigma_s({\bf k_0},
\{\alpha\}).
\label{opthe}
\end{equation}
The first term in the right side contains
$4\pi/\widehat k_0(q_0)$ instead of the known  $4\pi/k_0$  since  the
velocity of the falling flow  is $\hbar\widehat k_0(q_0)/m$.
The  $R^*(q_0)$ factor
takes into account a modification of the flow due to  coherent
interaction of the UCN with the matter medium.

The losses $\widetilde N_l(k_0,\{\alpha\})$ and
$\widetilde N_{il}(k_0,\{\alpha\})$ of UCN with given energy
$k_0$ are obtained by the integration of the losses
over
directions of UCNs in the trap.
Furthermore, $\widetilde N_{il}(k_0,\{\alpha\})=4\pi k_0
\widetilde\sigma_c(k_0,\{\alpha\})$,   where
$\widetilde\sigma_c(k_0,\{\alpha\})$
is the averaged cross section.
It commonly is that
an isotropic angular distribution of UCNs in the trap is
assumed \cite{serebrov,ignat}. In this case
\begin{equation}
\widetilde
\sigma_c(k_0,\{\alpha\})
=\frac{1}{4\pi}
\int_0^{2\pi}\,d\phi_0
\int_0^{\pi/2}\sigma_c({\bf k_0},
\{\alpha\})\cos\theta_0\sin\theta_0\,d\theta_0
=
\frac{1}{4\pi} \int_{q_0<k_0} \sigma_c({\bf k_0},
\{\alpha\})\, \frac{d^2q_0}{k_0^2}\,,
\label{avqu}
\end{equation}
where
angles $(\theta_0,\phi_0)$ specify  direction of
${\bf k_0}$.

Since interference effects from impurities are negligible (see Appendix
A and Introduction), all the above relations are applicable to
the interaction of UCN with  a single impurity.
For a spherical impurity
the cross sections $\sigma_c(k_0,\{\alpha\})$ and $\widetilde
\sigma_c(k_0,\{\alpha\})$  will be
denoted respectively as $\sigma_c({\bf k_0}, z_c,a)$ and
$\widetilde\sigma_c({\bf k_0}, z_c,a)$. As before, $z_c$ is a distance
from the center of the impurity to the border $z=0$, and $a$ is the
impurity radius.  The  macroscopic cross section is
obtained multiplying $\sigma_c({\bf k_0}, z_c,a)$ by the impurity
density which, generally, depends on $a$. Then it is useful
in addition to
introduce the macroscopic cross section for a
certain convenient density $n_0=10^{14}/{\rm
cm}^3$. This macroscopic cross section $\Sigma_{0c}({\bf
k_0},z_c,a)$ and the averaged cross section
$\widetilde\Sigma_{0c}(k_0,z_c,a)$ are given by
\begin{eqnarray}
\Sigma_{0c}({\bf k_0}, z_c,a)=n_0\sigma_c({\bf k_0}, z_c,a)\,,
\quad
n_0=10^{14}/{\rm cm}^3\,,\quad
\widetilde\Sigma_{0c}(k_0,z_c,a)=
n_0\widetilde\sigma_c(k_0,z_c,a)\,.
\label{stcs}
\end{eqnarray}
The coefficient of the UCN losses is the relation of the total
number of the lost neutrons to the total number of neutrons falling on
the boundary.  For $n=n_0$, the coefficient $\mu_0({\bf k_0},a)$ of the
UCN losses, and the averaged  coefficient $\widetilde\mu_0(k_0,a)$ are
given by
\begin{equation}
\mu_0({\bf k_0},a)=\int_a^\infty
\Sigma_{0c}({\bf k_0}, z_c,a) dz_c\,,\quad
\widetilde\mu_0(k_0,a)=\int_a^\infty
\widetilde\Sigma_{0c}(k_0,z_c,a) dz_c\,.
\label{losso}
\end{equation}
UCN losses arise only if the UCN interaction potential
has the imaginary part somewhere.  Indeed,
the current density  (\ref{current}) can be given
in  terms of the wave function $\Psi({\bf r})$ in the medium.
Off impurities  $\Psi({\bf r})$ coincides with
$\psi_2({\bf r},{\bf k_0},
\{\alpha\})$ discussed above.
This $\Psi({\bf r})$
tends to zero at $z\to\infty$ and obeys the
Schr\"{o}dinger equation as  follows:
\begin{equation}
\nabla^2
\Psi({\bf r})=\frac{2m}{\hbar}\Biggl( \widetilde V({\bf r})- E_0\Biggl)
\Psi({\bf r})\,,
\label{schrr}
\end{equation}
where the $\widetilde V({\bf
r})$  potential includes the impurity potential, too.
As is usually done, multiplying both parts of
(\ref{schrr}) by $\Psi^*(\overrightarrow{ l},z)$, subtracting from the obtained
equation its complex conjugate, and integrating the result over the
$z\geq0$ half-space, one obtains that
the total number $N_l({\bf k_0})$ of
neutrons lost per second is given by
\begin{equation}
N_l({\bf k_0})
= -\int Im\widetilde V({\bf r})\Biggl| \Psi({\bf
r})\Biggl|^2\,d^3r\,,
\label{curcon}
\end{equation}
where $j(\overrightarrow{ l})$
is the current density at $z=0$.
If $Im\widetilde V({\bf r})$ is zero all over (and therefore
UCN absorption is lacking), then the UCN losses are absent.

\section{UCN losses from  spherical impurity.}

The results of Sec. II are applied
in this Section to the
interaction of UCN with  a spherical impurity.
Hence $\widetilde
\psi_0({\bf r},\overrightarrow{ p},\kappa, \{\alpha\})$ is
given by (\ref{tfmat}).  Furthermore,
\begin{equation}
\frac{e^{-\kappa|{\bf r}
-{\bf r_c}}|}{|{\bf r}
-{\bf r_c}|}=\int
e^{i\overrightarrow{ q}\cdot
(\overrightarrow{ l}-\overrightarrow{ l_c})}
e^{-\widehat k(q)|z-z_c|}\frac{d^2q}{2\pi \widehat\kappa(q)}\,.
\label{outw}
\end{equation}
Thus $\widehat\psi(\overrightarrow{ q},\overrightarrow{ p},\kappa,
\{\alpha\})
\equiv
\widehat\psi(\overrightarrow{ q},\overrightarrow{ p},\kappa,
{\bf r_c},a)$ in eq.(\ref{tif1}) is represented as follows:
\begin{equation}
\widehat\psi(\overrightarrow{ q},\overrightarrow{ p},\kappa,
{\bf r_c},a)=
\frac{e^{-\widehat k(q)z_c}}{2\pi \widehat\kappa(q)}
e^{-i\overrightarrow{ q}\cdot
\overrightarrow{ l_c}}
e^{i\overrightarrow{ p}\cdot
\overrightarrow{ l_c}}e^{-\widehat k(p)z_c}
F(\kappa,a,\cos\vartheta(p,q))\,,
\label{wffs}
\end{equation}
where
\begin{equation}
\cos\vartheta(p,q)=-[\overrightarrow{
p}\cdot\overrightarrow{q}+\widehat\kappa(p)\widehat\kappa(q)]/\kappa^2\,,
\label{cosi}
\end{equation}
and other definitions are given in
(\ref{hatkkl}).
Furthermore,
$\widetilde B(\overrightarrow{ q},{\bf k_0},\{\alpha\})$
in (\ref{tif1})
is represented as follows:
\begin{equation}
\widetilde
B(\overrightarrow{ q},{\bf k_0},\{\alpha\}) = \frac{\widehat
k_0(q_0)e^{-i\overrightarrow{ q}\cdot\overrightarrow{ l_c}} e^{-\widehat
k(q)z_c}}{\pi \widehat\kappa(q) (\widehat k_0(q_0)+i\widehat\kappa(q_0))}
e^{i\overrightarrow{ q_0}\cdot
\overrightarrow{ l_c}}e^{-\widehat k(q_0)z_c}
\widetilde
F(\overrightarrow{ q},{\bf k_0},z_c,a)    \,,
\label{tildf}
\end{equation}
where $\widetilde
F(\overrightarrow{ q},{\bf k_0},z_c,a)$
is independent of $\overrightarrow{ l_c}$.
Using (\ref{tif1}), the equation for
$\widetilde
F(\overrightarrow{ q},{\bf k_0},z_c,a)$
is found to be
\begin{equation}
\widetilde
F(\overrightarrow{ q},{\bf k_0},z_c,a)=
F(\kappa,a,\cos\vartheta(q_0,q))
-
\int
\frac{e^{-2\widehat\kappa(p) z_c}
R(p)}{2\pi\widehat\kappa(p)}
\widetilde
F(\overrightarrow{ p},{\bf k_0},z_c,a)
F(\kappa,a,\cos\vartheta(p,q))
d^2p\,.
\label{relnul}
\end{equation}
By (\ref{capt}) and (\ref{stcs}), the macroscopic cross section
of the UCN losses
$\Sigma_{0c}({\bf k_0}, z_c,a)$
is as follows:
\begin{eqnarray}
\Sigma_{0c}({\bf k_0}, z_c,a)
= 16n_0\pi
\widehat k_0(q_0)\Biggl[Im\Biggl(R^*(q_0)
\frac{e^{-2\widehat\kappa(q_0)z_c}\widetilde
F(\overrightarrow{ q_0},{\bf k_0},z_c,a)}{[\widehat
k_0(q_0)+ i\widehat \kappa(q_0)]^2}\Biggl)
\nonumber\\
-2\int_{q<k_0}\frac{d^2q}{2\pi}
\biggl|\frac{e^{-2\widehat\kappa(q) z_c}}
{\widehat k_0(q)+i\widehat\kappa(q)}\biggl|^2\widehat
k_0(q) \biggl|\frac{e^{-2\widehat\kappa(q_0)z_c}\widetilde
F(\overrightarrow{ q},{\bf k_0},z_c,a)}{\widehat
k_0(q_0)+ i\widehat \kappa(q_0)}\biggl|^2\Biggl]\,.
\label{tphccrsl}
\end{eqnarray}
As was discussed in the Introduction, in this paper
UCN losses due to the UCN capture by
impurities  are mainly considered,
$ImU=0$ being kept. Then (\ref{tphccrsl}) is
turned out as follows:
\begin{eqnarray}
&&\hspace*{-0.5cm}\Sigma_{0c}({\bf k_0}, z_c,a)=
16n_0\pi \frac{\widehat k_0(q_0)}{\kappa_0^2}
e^{-2\widehat\kappa(q_0)z_c}
\nonumber\\
&&\hspace*{-0.5cm}\times
\Biggl[Im\widetilde
F(\overrightarrow{ q_0},{\bf k_0},z_c,a)
-2\int_{q<k_0}\frac{d^2q}{2\pi\kappa_0^2}e^{-2\widehat\kappa(q) z_c}\widehat
k_0(q)
\biggl|\widetilde
F(\overrightarrow{ q},{\bf k_0},z_c,a)\biggl|^2\Biggl]\,.
\label{phccrsl}
\end{eqnarray}
Eq.(\ref{relnul}) is easy solved  when only
the s-wave UCN-impurity interaction is taken into account.
Then the UCN-impurity scattering amplitude is independent of the
scattering angle, and so
$\widetilde F(\overrightarrow{ q},{\bf
k_0},z_c,a)$ is independent of the scattering angle, too.
The solution of eq.(\ref{relnul}) is
easy found (cf. appendix 6.13 in \cite{ignat}) being
as follows:
\begin{eqnarray}
\widetilde F(\overrightarrow{
q},{\bf k_0},z_c,a) \equiv \widetilde F_0(\kappa,z_c,a)
=
\frac{1}{F_0^{-1}(\kappa,a)- J(z_c,\kappa)}\,,
\nonumber\\
J(z_c,\kappa)=-\int
\frac{e^{-2\widehat\kappa(p) z_c} R(p)}{\widehat\kappa(p)}p\,dp\,.
\label{tilf}
\end{eqnarray}
In addition, eq.(\ref{relnul}) is analytically
solved for the large size impurity when $\kappa a>>1$.
Closed impurities being considered, then
$z_c/a\geq1$.  Due to the $\exp[-2\widehat\kappa(p)z_c]$ factor,
only $p\to0$ are  significant in the integral
in (\ref{relnul}). So $\widetilde F(\overrightarrow{
p},{\bf k_0},z_c,a)$ can be replaced by $\widetilde F(0,{\bf
k_0},z_c,a)$.  By taking $q=0$ and calculating the integral
over $p$ at $\kappa z_c>>1$, a simple equation for
$\widetilde
F(0,{\bf k_0},z_c,a)$
is obtained. In this case
$\widetilde
F(0,{\bf k_0},z_c,a)$
is as follows:
\begin{equation}
\widetilde F(0,{\bf
k_0},z_c,a)
=\frac{F(\kappa,a,-\widehat\kappa(q_0)/\kappa)}{1+\frac{e^{-2\kappa
z_c}}{2z_c}R(0)F(\kappa,a,-1)}\,,
\label{larrr}
\end{equation}
where
$R(0)$ is given by (\ref{reflco}) at $q=0$, and $F(\kappa,a,-1)$ is
the backward scattering amplitude. Substituting (\ref{larrr}) into
(\ref{relnul}), one easy derives
$\widetilde F(\overrightarrow{ q},{\bf k_0},z_c,a)$ for a general
$\overrightarrow{ q}$, but it will not be used below.

The foresaid
equations of this Section are easy expended to the case of an arbitrary
shape impurity. In this case
$a$ and $z_c$ are replaced by a set $\{\alpha\}$ of
parameters determining the shape, size and placement of the impurity
and $F(\kappa,a,\cos\vartheta(p,q))$ is replaced by a
relevant scattering amplitude.

As is usually done for the spherically symmetrical potential,
$F(\kappa,a,
\cos\theta)$
is represented
through
partial amplitudes $F_l(\kappa,a)$ as  follows:
\begin{equation}
F(\kappa,a,
\cos\theta)=\sum_{l=0}^{l=\infty}(2l+1)
F_l(\kappa,a)P_l(\cos\theta)\,,
\label{waves}
\end{equation}
where
$P_l(\cos\theta)$ is the Legender polynomial.
In this case
\begin{equation}
P_l(\cos\vartheta(p,q))=
\sum_{m=-l}^{m=l}e^{im(\phi_q-\phi_p)}
\frac{\Gamma(l-m+1)}{\Gamma(l+m+1)}
(-1)^l
P_l^m(\widehat\kappa(p)/\kappa)
P_l^m(\widehat\kappa(q)/\kappa)\,,
\label{plm}
\end{equation}
where $\phi_q$ is an azimuthal angle of $\overrightarrow{ q}$.
For $m>0, x>1$, one finds that
\begin{eqnarray}
&&\hspace*{-0.5cm}P_l^m(x)=(\sqrt{x^2-1})^m\frac{d^mP_l(x)}{dx^m}\,,
\quad
P_l^m(-x)=(-1)^l P_l^m(x)\,,
\nonumber\\
&&\hspace*{-0.5cm}
P_l^{-m}(x)=\frac{\Gamma(l-m+1)}{\Gamma(l+m+1)}P_l^m(x)\,.
\label{polmp}
\end{eqnarray}
To solve (\ref{relnul}) beyond  approximations (\ref{tilf}) and
(\ref{larrr}), one represents
$\widetilde F(\overrightarrow{ p},{\bf k_0},z_c,a)$
as
\begin{equation}
\widetilde
F(\overrightarrow{ q},{\bf k_0},z_c,a)=
\sum_{l=0}^{l=\infty}(2l+1)(-1)^l\sum_{m=-l}^{m=l}
e^{im(\phi_q-\phi_{q_0})}
\frac{\Gamma(l-m+1)}{\Gamma(l+m+1)}
\widetilde F_l^{(m)}(\kappa,q_0,z_c,a)
P_l^m(\widehat\kappa(q)/\kappa)\,.
\label{tilwv}
\end{equation}
From (\ref{relnul}), equation for
$\widetilde F_l^{(m)}(z_c,\kappa,q_0,a)$
is found to be
\begin{equation}
\widetilde
F_l^{(m)}(\kappa,q_0,z_c,a)=F_l(\kappa,a)P_l^m(\widehat\kappa(q_0)/\kappa)-
F_l(\kappa,a)
\sum_{l^\prime=|m|}^{\infty}C_{ll^\prime}^{(m)}(z_c,\kappa)
\widetilde F_{l^\prime}^{(m)}(\kappa,q_0,z_c,a)\,,
\label{eqlw}
\end{equation}
where
\begin{equation}
C_{ll^\prime}^{(m)}(z,\kappa)=-(-1)^{l^\prime}\widetilde C_{ll^\prime}^{(m)}(z,\kappa)
\frac{\Gamma(l^\prime-m+1)}{\Gamma(l^\prime+m+1)}(2l^\prime+1)\,,
\label{coe}
\end{equation}
\begin{equation}
\widetilde C_{ll^\prime}^{(m)}(z,\kappa)
=-
\int_0^\infty
\frac{e^{-2\widehat\kappa(q)z}}{\widehat\kappa(q)}qR(q)
P_l^m(\widehat\kappa(q)/\kappa)
P_{l^\prime}^m(\widehat\kappa(q)/\kappa)\,dq\,.
\label{coef}
\end{equation}
In this case $R(q)$ is given by (\ref{reflco}) where $\widehat
k_0(q)=i|\widehat k_0(q)|$ for $q>q_0$. Hence  $J(z,\kappa)$ in
(\ref{tilf}) is none other than $\widetilde C_{00}^{(0)}(z,\kappa)$.
Both (\ref{waves}) and (\ref{plm}) are below  approximated  by a
finite number $l\leq l_{max}$ of terms \footnote{For this reason  we do
not discuss a convergence of the above series.}.

For a constant potential, $F_l(\kappa,a)$ in
(\ref{waves}) is given by an analytical continuation in energy $E$ to
$E=U-\hbar\kappa^2/2m$ of the corresponding partial amplitude
\cite{landau} as follows:
\begin{equation}
F_l(\kappa,a)=(-1)^l\,\frac{\pi}{2}
\frac{\kappa I_{l+1/2}^\prime(\kappa a)J_{l+1/2}(\widetilde k a)-
\widetilde kI_{l+1/2}(\kappa a)J_{l+1/2}^\prime(\widetilde k a)}
{\widetilde k K_{l+1/2}(\kappa a)J_{l+1/2}^\prime(\widetilde k a)
-\kappa K_{l+1/2}^\prime(\kappa
a)J_{l+1/2}(\widetilde k a)}\,,
\label{inpasph}
\end{equation}
where $I_p(x)$, $K_p(x)$ and $J_p(x)$ are relevant Bessel functions;
for any function $f(x)$, it is defined  $f^\prime(x)=df(x)/dx$.
Furthermore,
\begin{eqnarray}
\widetilde
k^2=2m(U_d+iW+E)/\hbar^2=\kappa_0^2(\beta+iw)+k_0^2
= k^2+iw\kappa_0^2;
\nonumber\\
k^2=\kappa_0^2\beta+k_0^2\,,\qquad\beta=U_d/U'\,,
\qquad
w=W/U'\,,
\qquad U'=Re U\,,
\label{tik}
\end{eqnarray}
where $-(U_d+iW)$ is the potential of the impurity, and more
definitions are given in (\ref{hatkkl}). In this case $U_d>0$ and
$W>0$.
If $ImU=0$, then real
and imaginary parts of $\widetilde C_{ll^\prime}^{(m)}(z,\kappa)$  in
(\ref{coef}) are given by
\begin{eqnarray}
Re\widetilde C_{ll^\prime}^{(m)}(z,\kappa)\equiv
C_{1ll^\prime}^{(m)}(z,\kappa)
=\int_0^\infty dq\,q\,e^{-2\widehat\kappa(q)z}
\frac{\widehat\kappa^2(q)-
\widehat k_0^2(q)}{\widehat\kappa(q)\kappa_0^2}
P_l^m(\widehat\kappa(q)/\kappa)
P_{l^\prime}^m(\widehat\kappa(q)/\kappa)
\nonumber\\
-2\int_{k_0}^\infty
\frac{dq}{\kappa_0^2}\,q\,e^{-2\widehat\kappa(q)z}
\sqrt{q^2-k_0^2}
P_l^m(\widehat\kappa(q)/\kappa)
P_{l^\prime}^m(\widehat\kappa(q)/\kappa)\,,
\nonumber\\
Im\widetilde C_{ll^\prime}^{(m)}(z,\kappa)\equiv
\widetilde C_{2ll^\prime}^{(m)}(z,\kappa)
=2\int_0^{k_0}
\frac{dq}{\kappa_0^2}\,q\,e^{-2\widehat\kappa(q)z}\, \widehat
k_0(q)
P_l^m(\widehat\kappa(q)/\kappa)
P_{l^\prime}^m(\widehat\kappa(q)/\kappa)\,.
\label{cofe}
\end{eqnarray}
By using of (\ref{tilwv}), of (\ref{eqlw}) and
of (\ref{cofe}), eq.(\ref{phccrsl}) can be transformed
as  follows:
\begin{eqnarray}
\Sigma_{0c}({\bf k_0}, z_c,a)=n_0 \frac{16\pi \widehat
k_0(q_0)}{\kappa_0^2}\,e^{-2\widehat\kappa(q_0)z}
\sum_{l=0}^\infty [-Im
F_l^{-1}(\kappa)](-1)^l(2l+1)
\nonumber\\
\times
\sum_{m=-l}^{m=l}
\frac{\Gamma(l-m+1)}{\Gamma(l+m+1)}
\biggl|\widetilde
F_{l}^{(m)}(\kappa,q_0,z_c,a)\biggl|^2\,.
\label{flosdt}
\end{eqnarray}
To derive (\ref{flosdt}), one substitutes (\ref{tilwv})
to (\ref{phccrsl}).
Then the second term
in the right side of (\ref{phccrsl})
is represented as   follows:
\begin{eqnarray}
i\sum_{m=-\infty}^\infty\sum_{l=|m|}^\infty
\sum_{l^\prime=|m|}^\infty\biggl[\widetilde
C_{ll^\prime}^{(m)}(z,\kappa)(2l+1)\frac{\Gamma(l-m+1)}{\Gamma(l+m+1)}
-
\widetilde
C_{ll^\prime}^{(m)*}(z,\kappa)
(2l^\prime+1)\frac{\Gamma(l^\prime-m+1)}{\Gamma(l^\prime+m+1)}\biggl]
\nonumber\\
\times
\widetilde
F_{l^\prime}^{(m)}(\kappa,q_0,z_c,a)
\widetilde F_{l}^{(m)}(\kappa,q_0,z_c,a)\,.
\label{ddeer}
\end{eqnarray}
Using then eq.(\ref{eqlw})
and taking into account the first term in the right side of
(\ref{phccrsl}),
one
obtains (\ref{flosdt}).
If (\ref{eqlw}) is cut off by $l\leq l_{max}$, then
$\widetilde F_{l}^{(m)}(\kappa,q_0,z_c,a)$
is approximated as follows:
\begin{equation}
\widetilde
F_{l}^{(m)}(\kappa,q_0,z_c,a)=\sum_{l^\prime=|m|}^{l_{max}}
A_{ll^\prime}^{(m)}(\kappa,z_c,a) P_{l^\prime}^m(\widehat\kappa(q_0)/\kappa)\,,
\label{tflm}
\end{equation}
where the $A_{ll^\prime}^{(m)}(\kappa,z,a)$ matrix  obeys the equation:
\begin{equation}
A_{ll_1}^{(m)}(\kappa,z,a)=F_l(\kappa,a)\delta_{ll_1}-
F_l(\kappa,a)
\sum_{l^\prime=|m|}^{l_{max}}C_{ll^\prime}^{(m)}(z,\kappa,a)
A_{l^\prime l_1}^{(m)}(z, \kappa,a)\,,
\label{aeqlw}
\end{equation}
$\delta_{jj^\prime}$ being the Kronecker symbol. Then
\begin{eqnarray}
\widetilde\mu_0(k_0,a)=n_0 \frac{4\pi
}{k_0^2}\int_a^\infty d\, z\,
\sum_{l=0}^{l_{max}}[-Im
F_l^{-1}(\kappa,a)]
(-1)^l(2l+1)
\nonumber\\
\times
\sum_{m=-l}^{m=l}
\frac{\Gamma(l-m+1)}{\Gamma(l+m+1)}
\sum_{l_1=|m|}^{l_{max}}
\sum_{l_2=|m|}^{l_{max}}
\widetilde C_{2l_1l_2}^{(m)}(z,\kappa)
A_{ll_1}^{(m)}(z, \kappa,a)A_{ll_2}^{(m)*}(z, \kappa,a)\,,
\label{aflsdt}
\end{eqnarray}
where $\widetilde C_{2l_1l_2}^{(m)}(z,\kappa)$ is defined in
(\ref{cofe}), and the right-top star denotes  complex conjugation.

Throughout the paper, parameters in
(\ref{hatkkl}) and in (\ref{tik}) will be taken
for the case of the UCN capture by ice
impurities in the beryllium trap.
As is commonly done, the optical potential
of a neutron is
$2\pi\hbar^2\rho_na_{sc}/m$ where $\rho_n$ is the nuclear density of
matter and $a_{sc}$ is the UCN scattering length.
The Be nuclear density is  $1.235\times
10^{23}$/cm$^3$, and the ice molecular density is
$0.3175\times10^{23}$/cm$^3$.  The $n -Be$, $n - O$ and
$n - p$ coherent scattering length is, respectively,  7.79 fm,
2.9 fm and -1.87 fm. The imaginary part of the $n-p$ scattering length
is $-4.63\times10^{-5}$ fm.  This is calculated from a neutron
absorption cross section for the neutron velocity of 2200 m/sec. The
ice potential is the sum of the $n-O$ and $n-H_2$ potentials.  The
$n-Be$ potential is found to be $U=2.505\times10^{-7}$ eV or in cm:
$L_0= 244.3$ cm. The absorption of UCN by Be
is neglected ($ImU=0$).
Then the parameters in (\ref{hatkkl}) and in
(\ref{tik}) are found to be
\begin{equation}
\kappa_0=1.1007\times10^{6}/\text{cm};\qquad \beta
=U_d/U=0.02452;\qquad
w= W/U=3.176\times10^{-6}\,.
\label{numd}
\end{equation}

\section{Small and large impurities}

To clarify main peculiarities of the UCN losses,
the limiting cases $\kappa a>>1$ and
$\kappa a<<1$ are discussed in this Section.

If higher wave  resonances ($l\geq1$) lie outside of the UCN energy
range, then, as  has been noted already and will be discussed below,
eq.(\ref{tilf}) is reasonable.
When, in addition, absorption of UCN by the matter is negligible,
then from (\ref{phccrsl})
the averaged coefficient $\widetilde\mu_0(k_0,a)$ of UCN
losses (\ref{losso})
for the impurity density $n_0$
is found to be
\begin{equation}
\widetilde\mu_0(k_0,a)
=-4\pi\,n_0
\int_a^\infty
\frac{J_2(z_c,\kappa)Im
F_0^{-1}(\kappa)}{k_0^2
|F_0^{-1}(\kappa,a)- J(z_c,\kappa)|^2}da \,,\quad
J(z_c,\kappa)\equiv \widetilde
C_{100}^{(0)}(z_c,\kappa)\,,
\label{avmu0}
\end{equation}
where $F_0(\kappa,a)$ is given by (\ref{inpasph}) at $l=0$ and
$\widetilde
C_{100}^{(0)}(z_c,\kappa)$ is defined by (\ref{cofe}).
In  accordance with  first principles $ImF_0^{-1}(\kappa,a)$ is
negative.
It follows  from (\ref{cofe})  that
$ImJ_2((z_c,\kappa))$
is negative, too. Thus to calculate (\ref{avmu0}) in the
leading approximation
at $W\to0$, only the leading  term
in $Re F_0^{-1}(\kappa,a)$ and in $Im F_0^{-1}(\kappa,a)$  are
important. Thus $Re F_0^{-1}(\kappa,a)$  is given by
 (\ref{inpasph}) at $l=0$  and $W=0$, while
$Im F_0^{-1}(\kappa,a)$ is approximated
as follows:
\begin{equation}
Im F_0^{-1}(\kappa,a)=w\frac{\kappa_0^2}{2k}\,\frac{\kappa^2(\cot
ka-ka/\sin^2ka) }{[\kappa\cosh(\kappa a) - k\cot
ka\sinh(\kappa a)]^2 }\,.
\label{swa}
\end{equation}
If the resonance does not occur, then rescattering can be neglected
and therefore
\begin{equation}
\widetilde\mu_0(k_0,a)\approx
\, n_0\frac{4\pi}{k_0^2}\int_a^\infty d\,
z_c\,J_2(z_c,\kappa)ImF_0(\kappa,a)\,.
\label{must0s}
\end{equation}
If in addition $\kappa_0a<<1$,
then from (\ref{swa}) it follows that
\begin{equation}
Im F_0(\kappa,a)=wa^3\kappa_0^2/3\,.
\label{zerf}
\end{equation}
Then, by using (\ref{must0s}), one obtains that
\begin{equation}
\widetilde\mu_0(k_0,a)
\approx\,
\frac{2\pi}{3k_0^2\kappa_0^2}n_0 w
a^3
\biggl[\kappa_0^2\arcsin(k_0/\kappa_0)
-k_0\sqrt{\kappa_0^2-k_0^2}\,\biggl]\,.
\label{zerra}
\end{equation}
Replacing $n_0$ in (\ref{zerra}) by  the true impurity density and
integrating the obtained expression over $a$, one comes
the result multiplying by 4 to eq.(\ref{intro}) with $\eta= p_Vw$ as
was announced in the Introduction \footnote{Multiplying by 4 is made
because the averaged cross section (\ref{avqu}) contains the 1/4
multiplication factor with respect to the average cross section defined
in \cite{serebrov,ignat}.}. The UCN losses are small in this case.  Much
larger losses arise when there is a resonance in the UCN-impurity
scattering amplitude as  is shown below using the
$\kappa a>>1$ approximation.

Only for simplicity one can assume in addition that
$k_0a>>1$ and $k_0^2a/\kappa>>1$. Then, by
using (\ref{larrr}) in the $ImU=0$ approximation and
after
integration over $\overrightarrow{ q_0}$
the averaged coefficient (\ref{avqu})
of the UCN losses
is found to be
\begin{equation}
\widetilde\mu_0(k_0,a)
\approx
n_0\,\frac{8\pi\,\kappa}{k_0\kappa_0^2}
\int_a^\infty \frac{d\,
z_c}{2z_c}\,\frac{e^{-2\kappa z_c}ImF(\kappa,a,-1)}{
|1+\frac{e^{-2\kappa
z_c}}{2z_c}R(0)F(\kappa,a,-1)|^2}\,.
\label{lamusts}
\end{equation}
Furthermore,
$d(e^{-2\kappa z_c}/(2z_c))\approx -(2\kappa e^{-2\kappa
z_c}/(2z_c))\,dz_c$  under the considered conditions.
Thus the integral over $z_c$ gives:
\begin{eqnarray}
\widetilde\mu_0(k_0,a)\approx
\frac{4\pi n_0\,ImF(\kappa,a,-1)}{\kappa_0^2k_0|
F(\kappa,-a,1)\sin\xi|}\Phi(\kappa,a)\,,
\nonumber\\
\cos\xi=Re\frac{F(\kappa,a,-1)R(0)}{|F(\kappa,a,-1)|}\,,
\label{arpssy} \\
\Phi(\kappa,a)=
\arctan\biggl(\frac{
e^{-2\kappa
a}|F(\kappa,a,-1)|}{(2a)|\sin\xi|}+
\frac{\cos\xi}{|\sin\xi|}\biggl)
-\arctan\frac{\cos\xi}{|\sin\xi|}
\,.
\label{arapssy}
\end{eqnarray}
The backward amplitude $F(\kappa,a,-1)$ is calculated by
(\ref{waves}) through
partial amplitudes (\ref{inpasph}), each
being approximated at $\kappa a>>1$, as  follows:
\begin{equation}
\kappa F_l(\kappa,a)\approx (-1)^l e^{2\kappa
a-\alpha_l}
\frac{\kappa-\widetilde k\cot(\widetilde ka-\pi
l/2-\alpha_l/2)}{\kappa+\widetilde k\cot(\widetilde ka-\pi
l/2-\alpha_l/2)}\,,\quad
\alpha_l=-\frac{(l+1/2)^2}{\kappa a}\,,
\label{asfb}
\end{equation}
where $l^2/\kappa a\la1$, $k a>>1$ and $\kappa a>>1$.
Keeping $\alpha_l$  is reasonable only if $l>>1$.
The resonance  is determined by the
condition
\begin{equation}
\kappa+ k\cot(ka-\pi
l/2-\alpha_l/2)=0\,,
\label{conres}
\end{equation}
where $k$ is $\widetilde k$ at $W=0$, cf. (\ref{tik}).
Therefore, nearby the resonance where $a|k_0^2-k_r^2|/k<<1$,
the backward amplitude  $F(\kappa,a,-1)$
is given by
\begin{equation}
F(\kappa,a,-1)=
-\frac{4 (2l+1)k_r^2(1+\beta) e^{2\kappa_r a} \,e^{-\alpha_l}}{
a(\kappa_r^2+\beta k_r^2)[(k_0^2-k_r^2)+ i\,w\kappa_0^2]}\,+\ldots\,,
\label{appam}
\end{equation}
where $k_r^2=2mE_r/\hbar^2$, $\kappa^2_r=\kappa_0^2-k_r^2$, and  $E_r$
is the resonance energy. The background is denoted by ellipsis.  Other
definitions are given in (\ref{hatkkl}) and (\ref{tik}).  Then the
resonance contribution to the  coefficient of the losses (\ref{losso})
is found to be
\begin{equation}
\widetilde\mu_0(k_0,a)\approx
\frac{4\pi n_0\,w\Phi(\kappa,a)}{k_r|\sin\xi|
\sqrt{(k_0^2-k_r^2)^2+(w\kappa_0^2)^2}}\,,
\label{assy}
\end{equation}
where $\Phi(\kappa,a)$ is calculated by (\ref{arapssy}) with
the understanding that the whole $F(\kappa,a,-1)$ is substituted
into (\ref{arapssy}), its background being included too.
If $\kappa a\to\infty$, then mainly the $l^2\sim \kappa a$ terms
contribute to the background. So $F(\kappa,a,-1)\sim a\exp(2\kappa a)$.
Furthermore, the imaginary part of the
partial amplitude (\ref{asfb}) is given by its derivative with
respect to $W$. Hence $ImF(\kappa,a,-1)/|ReF(\kappa,a,-1)|\sim
(w\kappa_0^2a/k)$, where $w$ and $k$ are defined in (\ref{tik}). So the
resonance losses (\ref{assy}) prevail the background ones when
$a^2|k_0^2-k_r^2|<<1$. It follows from
(\ref{conres}) that under the discussed conditions
many resonances arise, each being approximately determined by
the condition $k a\approx2\pi s$ provided $2\pi s/(ka)<<1$, and where
$s$ is an integer  including zero.  At  given $l$ the distance
between neighboring resonances is $2\pi k/a$ which is much greater than
$l/a^2$.  Resonances for $l$ and $l+4n$ where $n$ is an integer number,
are close to each other as $\sim l/a^2$.  Hence
the resonances do not overlap in the  region
$a^2|k_0^2-k_r^2|<<1$, and so the total
resonance contribution to $\widetilde\mu_0(k_0,a)$ is a sum over
resonances.

The discussed asymptotics is not fully achieved
for   $a \leq880$ \AA$\,$ considered below.
In this case a
single $s$-wave resonance occurs in the UCN-impurity scattering
amplitude, and a single $p$-wave resonance is added if $a\geq740$ \AA.
Then a nonresonance piece of $\exp(-2\kappa a)F(\kappa,a,-1)$ is rather
$\sim\kappa^{-1}$ than $\sim a$.  Hence  in the  $a^2|k_0^2-k_r^2|>>1$
range both terms on the right side of (\ref{arapssy}) cancel each other
and therefore $\Phi(\kappa,a)$ decreases.  In this case  rescatterings
cease to be important.  The resonance losses (\ref{assy}) prevail the
background ones when $a|k_0^2-k_r^2|<<k_r$.

In spite of the fact that
rescatterings from the wall, generally, spread
the resonance,
the width of the  resonance (\ref{assy}) remains  $\sim
w\kappa_0^2$.  This is due to
the fact that the main contribution to (\ref{lamusts}) arises from so
great $z_c$ that $\Gamma_2(z_c)\sim\max[|k_0^2-k_r^2|,w\kappa_0^2]$.
Indeed, approximating the
integrand
in (\ref{lamusts}) by its resonance piece, one obtains that
\begin{eqnarray}
&&\hspace*{-0.8cm}
\widetilde\mu_0(k_0,a)
\approx n_0\, \int_a^\infty
\frac{4\pi\Gamma_1\Gamma_2(z_c)}{
k_r^2[(k_0^2-k_r^2+\epsilon(z_c))^2+(\Gamma_1+\Gamma_2(z_c))^2]}\,d\,
z_c\,,
\nonumber\\
&&\hspace{-0.5cm}
\epsilon(z_c)=\frac{4(2l+1)k_r^2(1+\beta)(\kappa_0^2-2k_r^2)
e^{2\kappa_r(a-z_c)}}{2z_ca\kappa_0^2(\kappa_0^2+\beta k_r^2)}\,,
\nonumber\\
&&\hspace*{-0.8cm}\Gamma_1=w\kappa_0^2\quad
\Gamma_2(z_c)=\frac{8(2l+1)k_r^3(1+\beta)\kappa
e^{2\kappa_r(a-z_c)}}{2z_ca\kappa_0^2(\kappa_0^2+\beta k_r^2)}\,.
\label{murez}
\end{eqnarray}
The expression under the integral is non other than the
resonance piece of the UCN capture cross section
$\sigma_c(k_0,z_c,a)$ introduced in Sec. II.
For simplicity, $l\sim1$ is considered. The integral diverges
exponentially up to so great $z_c$ that
$\Gamma_2(z_c)\sim\max[|k_0^2-k_r^2|,w\kappa_0^2]$.  Hence these
$z_c$  give the main contribution  to the integral as was announced
above.

The shape of the resonance peak assigned to (\ref{assy})
is quite different from the shape of the Breit-Wigner  resonance.
Indeed, (\ref{assy}) contains
$1/\sqrt{(k_0^2-k_r^2)^2+(w\kappa_0^2)^2}$, but not
$1/[(k_0^2-k_r^2)^2+(w\kappa_0^2)^2]$. Besides,
$\Phi(\kappa,a)$ is significantly altered
when $k_0^2$ runs  within the resonance interval. In
particular, $\Phi(\kappa,a)$  is asymmetric with respect to the top of
the resonance.  The main features of $\widetilde\mu_0(k_0,a)$
are demonstrated by Fig.1 where $\widetilde\mu_0(k_0,a)$ for
$a\kappa_0=6.5$ is shown
against the UCN reduced energy $y=E/U=k_0^2/\kappa_0^2$.
Its
detailed behaviour in the resonance range is shown, too (the
right-side figure).
\begin{figure}[t] \begin{center}
\includegraphics[width=8.6cm,
height=6.5cm]{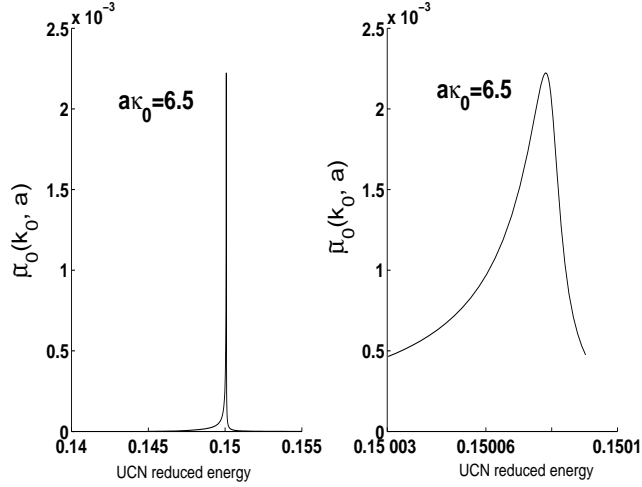} \end{center}\caption{\label{fig1}Coefficient
$\widetilde\mu_0(k_0,a)$ of UCN losses at $a\kappa_0=6.5$ against the
UCN reduced energy $k_0^2/\kappa_0^2$.}
\end{figure}

UCN losses per second
are determined by the integral over $k_0^2$ of the expression which
is $\widetilde\mu_0(k_0,a)$ multiplied by
the magnitude of the UCN flux ${\cal F}(k_0^2)$, see Sec. VI for
more details.  The leading contribution $\tau_r^{-1}$ to the integral
gives the range: $w\kappa_0^2<<|k_0^2-k_r^2|<<1/a^2$.  Since
${\cal F}(k_0^2)$ is an approximately constant in the above interval,
$\tau_r^{-1}$ is given by
\begin{equation}
\tau_r^{-1}\approx
{\cal F}(k_r^2) \frac{2\pi^2n_0}{\kappa_r k_r^2}\,w
\ln[1/w(\kappa_0a)^{2}]\,,
\label{mura}
\end{equation}
This expression
can be also  derived directly from (\ref{murez}). Indeed, the
integration of (\ref{murez}) over $k_0^2$ gives the result:
$16\pi^2\Gamma_1\Gamma_2(z_c)/(\Gamma_2(z_c)+\Gamma_1)$.  This
expression is integrated over $z_c$ by means of introducing
$\Gamma_2(z_c)$ as
the integration variable, and eq.(\ref{mura}) arises.

The relation of the resonance losses to the background
ones roughly is $\ln[1/w(\kappa_0a)^{2}]$ which is
$\sim10$ under the conditions considered.
Generally, the resonance in the partial amplitude (\ref{asfb}) increases
an average nonresonance magnitude of $ImF_l(\kappa,a)$ over the
integration interval that increases the background piece of
the UCN losses.  If the resonance does not occur, then the partial
amplitude (\ref{asfb}) has a zero (in the $W=0$ limit) in the
integration interval.  It, generally, reduces
$ImF_l(\kappa,a)$ that decreases the losses.  As a result, partial
waves of the last type contribute to the losses   about  a percent or
smaller.  These rough estimations are confirmed by the numerical
results of Sec. VI.  An accuracy of (\ref{arpssy}) has been estimated
to be about (20--30)$\%$.

It is instructive to compare
(\ref{mura}) with the losses from  the scattering of
UCN by a vacuum cavity \cite{ignat}. By contrast to foresaid, in the
last case $U_d=W=0$ in (\ref{tik}) , but the imaginary part $U''=\eta
U'$ of the $U$ potential is taken into account.
Examining (\ref{tphccrsl}) with $U_d=W=0$ and at $a\kappa\to\infty$,
one concludes that
the
resonance piece of the UCN losses is again given by (\ref{murez}) with
the understanding that $\Gamma_1$ is $\Gamma_1=\eta
k_r^2/(a\kappa_r)$ and, in addition, that $\Gamma_2(z_c)$ and
$\epsilon(z_c)$  are calculated by (\ref{asfb}) at $\beta=0$.
For $l=0$ it
agrees with the expression (17) of
appendix 6.13 in \cite{ignat} taking into account the difference in the
definition of
the average cross section
and up to certain inaccuracies \footnote{
In \cite{ignat}, there is a mistakable
extra factor 2 in $\Gamma_1$ and certain
insignificant inaccuracies
in  $\epsilon$ and $\Gamma_2$.}
in \cite{ignat}.
Instead of (\ref{mura}),  the UCN  losses $\tau_{cv}^{-1}$ per second
are now as follows:
\begin{equation}
\tau_{cv}^{-1}\approx {\cal F}(k_r^2) \frac{2\pi^2n_0}{
\kappa_r^2\kappa_0^2 a} \eta\log[1/\eta(\kappa_0a)]\,.
\label{murak}
\end{equation}
The relation of (\ref{mura} to (\ref{murak}) is mainly as
was announced in the Introduction \footnote {The losses explicitly
depend on the cavity radius $a$ as $1/a$, but not $a^3$
seemingly occurring in eq.(20) of appendix 6.13 in
\cite{ignat}}.

\section{Possible increasing of UCN losses for non-spherical
impurities}

The leading contribution to the UCN losses at large $a$ is given by the
sum over  resonances of fractional losses (\ref{mura}).
If the resonance is degenerated in some quantum number, then
it is
represented by a single term in the sum.
Indeed, the numerator and the
denominator in (\ref{assy}) each is proportional to the number of the
degenerated resonances.   In particular, each a resonance in the $l$-wave
partial amplitude   is $(2l+1)$-degenerated with respect
to the azimuthal quantum number, but its contribution (\ref{assy}) to
$\widetilde\mu_0(k_0,a)$ contains no $(2l+1)$-multiplier. A spherical
shape distortion is able to split the resonance into  the $(2l+1)$
resonances. Then the UCN losses due to this resonance could increase
by about $(2l+1)$ times. Nevertheless,  the volume occupied by
impurity could remain about the same.  It is
demonstrated below with an example of a small quadruple distortion of
the spherical impurity when
the shape of the impurity is given by
\begin{equation}
r=a+\frac{1}{2}\varepsilon_{ps}n_pn_s,\quad\varepsilon_{ss}=0,\quad
\varepsilon_{ps}=\varepsilon_{sp},
n_s=r_s/r,
\quad n_s^2=1,\quad
\varepsilon_{rs}^2<<a^2\,,
\label{shape}
\end{equation}
where the $\varepsilon_{ps}$ set gives the shape distortion.
The summation over twice-repeated indices is performed.  In this
case $ r_s$ is the $s$-component of the radius-vector ${\bf r}$, and
the center of the spherical impurity is placed in the point of origin.
For the sake of simplicity, only    mixing of states within the $l=1$
multiplet is considered. The wave vector of the falling wave is
kept to be $(0,0,i\kappa)$. If $a\kappa_0>>1$, then the $l=1$ piece
$\psi_{mat}({\bf r})$
of the UCN wave function
outside the impurity, and its
derivative $\psi_{mat}^\prime({\bf r})$ with respect to $r$,
are given by
\begin{equation}
\psi_{mat}({\bf r})\approx-\frac{3 e^{\kappa
r}}{2\kappa r}n_3 +{\bf fn}\frac{e^{-\kappa r}}{r}\,,\quad
\psi_{mat}^\prime({\bf r})\approx-\frac{3e^{\kappa r}}{2r}n_3
-{\bf fn}\frac{e^{-\kappa r}\kappa}{r}\,.
\label{defmat}
\end{equation}
Only the exponentially large term is
kept in  the falling  wave in (\ref{defmat})
for the calculation of the  resonance piece of the scattering amplitude
${\bf fn}$ in the leading approximation.
The $l=1$
piece
$\psi_{imp}({\bf r})$
of the UCN wave function inside the impurity
nearby  the boundary (\ref{shape}),
and its derivative
$\psi_{imp}^\prime({\bf r})$ with respect to $r$ are given by
\begin{equation}
\psi_{imp}({\bf r})\approx\frac{\cos\widetilde kr}{r} {\bf
dn}\,,\quad
\psi_{imp}^\prime({\bf r})\approx-\frac{\widetilde k\sin\widetilde
kr}{r} {\bf dn}\,,
\label{defim}
\end{equation}
where ${\bf
d}=(d_1\,,d_2\,,d_3)$ is an independent of ${\bf r}$.

At
$\varepsilon_{rs}=0$, the  resonance magnitude $k_r$ of the wave vector
is found from eq.(\ref{conres}) at $l=1$
(and  $\alpha_1=0$).
To find the leading term in the backward
amplitude, the linear in $\varepsilon_{rs}$ terms on the boundary
(\ref{shape})
need to be
kept only in (\ref{defim}).  Then equations
matching the wave functions
(\ref{defmat}) and (\ref{defim}) on the impurity boundary
(\ref{shape}), are as follows ($r=1,\,2,\,3$):
\begin{eqnarray}
-\frac{3 e^{\kappa
a}}{2\kappa a}\delta_{r3} =\frac{e^{-\kappa a}}{a}\,f_r
+\frac{\cos\widetilde ka}{a}\,d_r- \frac{\widetilde k\sin\widetilde
ka}{5a}\varepsilon_{rs}d_s\,,
\nonumber\\
\frac{3 e^{\kappa
a}}{2a}\delta_{r3}=\frac{e^{-\kappa a} \kappa}{a}\,f_r+ \frac{\widetilde
k\sin\widetilde ka}{a}d_r+\frac{\widetilde k^2\cos\widetilde
ka}{5a}\varepsilon_{rs}d_s\,,
\label{defeq}
\end{eqnarray}
where
$\delta_{rs}$ is the Kronecker symbol, and the summation
over twice-repeated indices is implied.  From (\ref{defeq}), the
backward amplitude $F(\kappa,a,-1)\approx-f_3$
about the resonance (\ref{conres}) is found to
be
\begin{equation}
F(\kappa,a,-1)
\approx
-\frac{12k_r^2(1+\beta) e^{2\kappa_r a}
\det_{33}[\widetilde x\delta_{js}+2k_r^2\varepsilon_{js}/(5a)]}{
a(\kappa_r^2+\beta
k_r^2)\det[(\widetilde
x+i\,w\kappa_0^2)\delta_{js}+2k_r^2\varepsilon_{js}/(5a)]}\,,
\widetilde x=k_0^2-k_r^2 \,,
\label{appamd}
\end{equation}
where $\det[\widetilde x\delta_{js}+2k_r\varepsilon_{js}/5]$ is the
determinant of the matrix whose elements are given in the square
brackets ($j,s=1\,,2\,,3$) of the above expression, and
$\det_{33}[\widehat x\delta_{js}+2k_r\varepsilon_{js}/5]$ is
(33)-minor of the determinant.  To obtain the coefficient of the
UCN losses,
the amplitude (\ref{appamd}) is substituted into (\ref{arpssy}). One
can see that (\ref{appamd}) has the resonance, if
$\widetilde x=x_j=-2k_r^2\varepsilon_j/(5a)$ where $\varepsilon_j$ is an
eigenvalue of the $\{\varepsilon_{js}\}$ matrix.
If  $a^2|x_j-x_s|>>1$  for any discussed $x_j$ and
$x_s$, then the UCN losses will be the sum over the
losses from every resonance, they will be three time more than the UCN
losses in the $\varepsilon_{rs}=0$ case.  The volume $V_{im}$ occupied
by impurity increases as follows:
\begin{equation}
V_{im}=\int
d\cos\theta
d\phi\,\frac{4\pi}{3}\biggl(a+\frac{1}{2}\varepsilon_{rs}n_rn_s
\biggl)^3
\approx
\frac{4}{3}\pi
a^3\biggl(1+\sum_{i=1}^3\frac{5x_i^2}{8a^2k_r^2}\biggl)\,.
\label{vol}
\end{equation}
Hence the relative change of the impurity volume
goes to zero when $a\to\infty$.
So, in the $a\to\infty$ limit, the distortion of the impurity
shape can increase the UCN losses remaining the impurity volume being
about the same.  As of now, it has not been studied whether the
discussed increasing of the UCN losses takes place in a realistic
range of $a$.

\section{Probability of UCN losses}

In this Section, there is given  UCN losses from
ice, spherical impurities calculated under conditions where the losses
have been measured experimentally.  An example of a radius impurity
distribution  is proposed which fits  the experimental losses of
UCNs in the beryllium trap. As it was noted in the Introduction,  now
experimental data are insufficient to fit them in the unique way.  The
proposed fitting must be considered only as an preliminary example.
\begin{table}[h] \begin{center}
\begin{tabular}{|c|c|c|c|c|c|c|}\hline $h_0$& 58 & 52 & 46 & 38 & 23 \\
\cline{1-6} $\gamma(h_0)$ & 0.4867& 0.4608&0.4334&0.3940&0.3065 \\
\cline{1-6} $a_r(1,h_0)$&5.109&5.424&5.791& 6.391&8.120\\ \cline{1-6}
$a_r(2,h_0)$&7.758&8.206&8.728&
9.583&12.05\\ \cline{1-6}
\end{tabular}
\end{center}
\caption{Reduced wave vector $\gamma(h_0)$
and $a_r(l,h_0)$ impurity radius.}
\end{table}
\begin{figure}[t]
\begin{center}
\includegraphics[width=8.6cm, height=7cm]{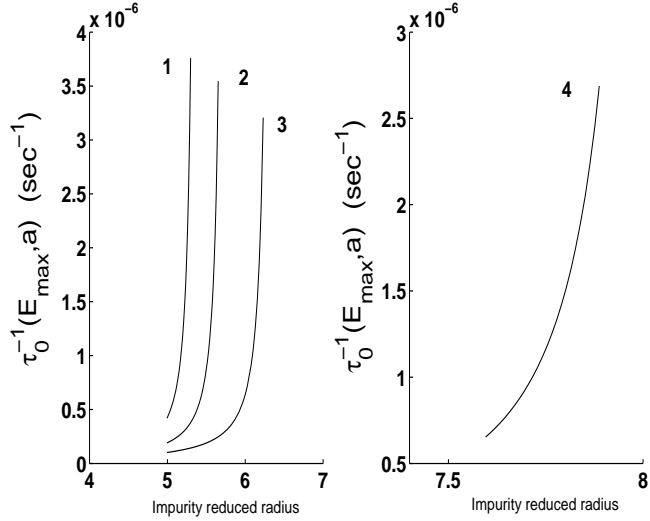}
\end{center}
\caption{\label{fig2}
Probability $\tau_0^{-1}(E_{\rm max},a)$ of UCN losses
against the impurity reduced radius in the region where
resonances are not available; the trap high is 52 cm (curve 1),
46 cm (curve 2), 38 cm (curve 3) and 23 cm (curve 4).}
\end{figure}
The  calculations are performed for a narrow
cylindrical beryllium trap \cite{serebrov}, its radius $R$ being $R=38$
cm, and its length $L$ being $L=14$ cm.  Numerical data
(\ref{numd}) are imployed.

For each of five discharges \cite{serebrov} the height $h_0$
of the trap is  respectively  58 cm, 52 cm, 46 cm,
38 cm and 23 cm. Simultaneously, $h_0$ measures $E_{\rm max}$ in
centimeters. The reduced wave vector
$k_{max}/\kappa_0$ assigned to $E_{\rm max}$,
will be denoted as $\gamma(h_0)$
where $h_0$ is measured by centimeters. As an example,
$\gamma(58)$ is the reduced wave vector assigned to
$h_0= 58$ cm. The reduced impurity radius $a\kappa_0$ is kept
within the range $4.994\leq a\kappa_0\leq9.633$ where the low limit
corresponds to the $s$-resonance
being at $\gamma=\gamma(58)+0.01= 0.4967$,
and the top limit corresponds to the $d$-resonance at
$\gamma=\gamma(63)+0.01=0.5173$.
An $l$-wave resonance is occurs at
$E=E_{\rm max}$ at a certain reduced radius of the impurity. This
reduced radius will be denoted as $a_r (l,h_0)$ where $h_0$ is
measured by centimeters.  The reduced radii in the region of interest
are given in Tab.1.

Experimentalists \cite{serebrov} determine a probability
of the
losses per second  $\tau^{-1}(E_{\rm max})$
of UCNs with energies up
to the given maximal energy $E_{\rm max}$.
In line with the foregone
text, it is useful to introduce in addition   $\tau_0^{-1}(E_{\rm
max},a)$, which is probability of UCN losses per second caused
\begin{figure*}[t]
\includegraphics[width=14cm, height= 8cm]{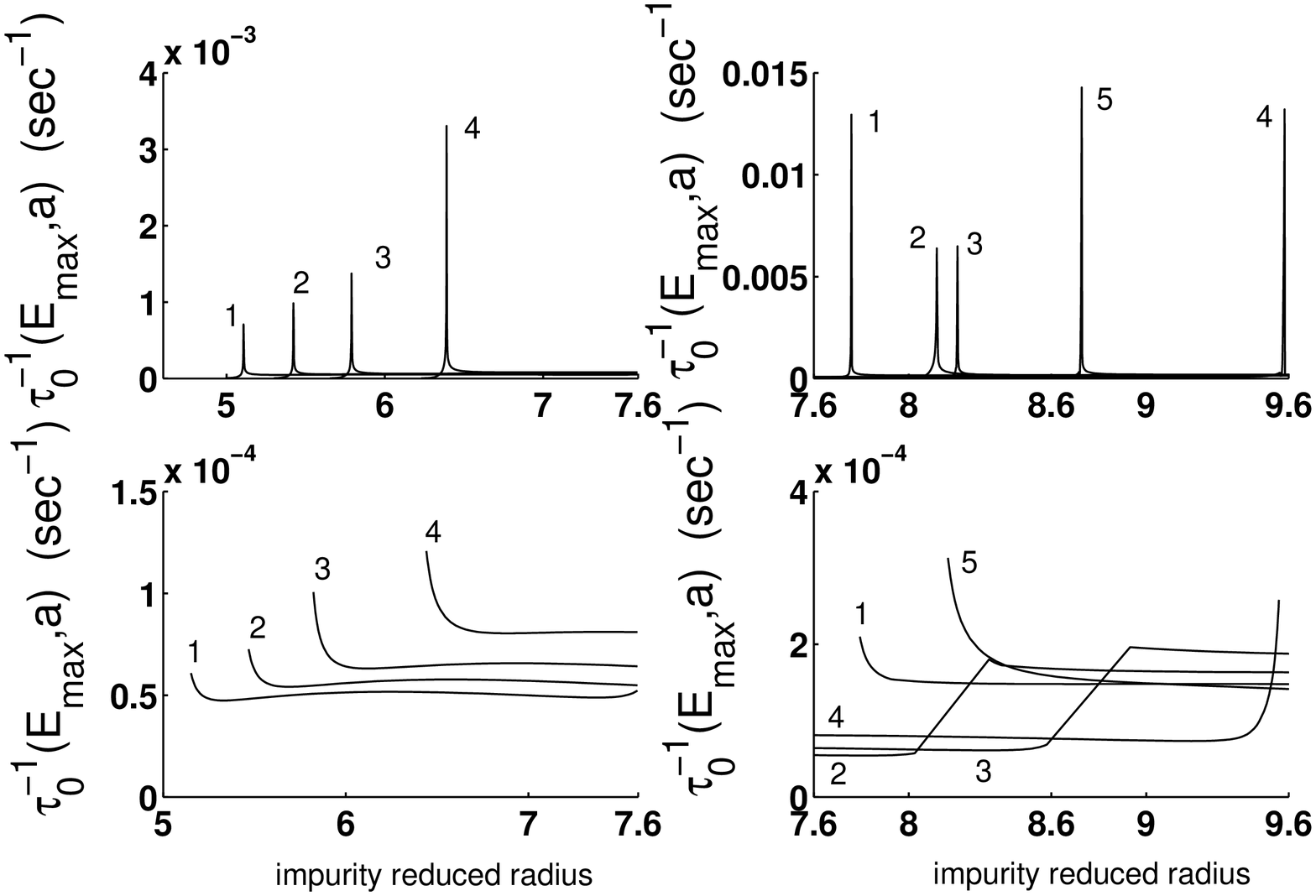}
\caption{\label{fig3}
UCN losses $\tau_0^{-1}(E_{\rm max},a)$ versus
the impurity reduced radius $a\kappa_0$
for $4.99\leq a\kappa_0\leq7.6$ where only $s$-resonance is available
and for $7.6\leq a\kappa_0\leq9.6$ where $s$- and $p$-resonances are
available;
the trap height is
58 cm (curves 1), 52 cm (curves 2), 46 cm (curves 3),
38 cm (curves 4) and 23 cm (curves 5).}
\end{figure*}
\begin{figure*}[t]
\includegraphics[width=14cm, height=11cm]{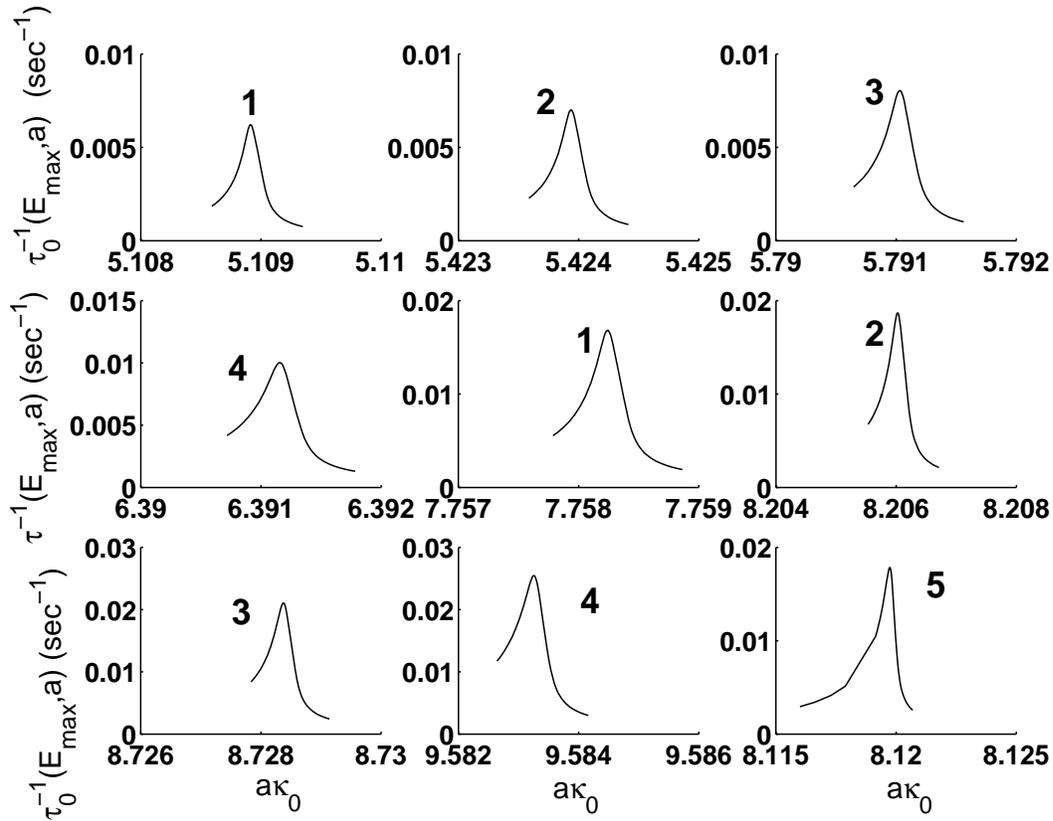}
\caption{\label{fig4}
UCN losses $\tau_0^{-1}(E_{\rm max},a)$ versus $a\kappa_0$
in the region of the picks;
the trap high is 58 cm (curves 1), 52 cm (curves
2), 46 cm (curves 3), 38 cm (curves 4) and 23 cm (curves
5). }
\end{figure*}

by the UCN capture by impurities of the given radius $a$ and of the
density  $n_0=10^{14}$/cm$^3$.  Hence $\tau_0^{-1}
(E_{\rm max},a)$
and $\tau^{-1}(E_{\rm max})$
are given by
\begin{eqnarray}
\tau_0^{-1}(E_{\rm
max},a)
=\frac{1}{N_1(E_{\rm max})}\int \rho(k_0,h)\frac{\hbar
k_0}{m}\widetilde\mu_0(E,a)\,k_0dk_0\,dS\,,
\label{tau0}\\
\tau^{-1}(E_{\rm max})=\int_0^\infty
n_0\frac{dn(a)}{da} \tau_0^{-1}(E_{\rm max},a) da\,,
\label{tau}
\end{eqnarray}
where $\widetilde\mu_0(E,a\kappa_0)$ is an averaged coefficient
of the UCN losses (\ref{losso}),
$n_0dn(a)$ is the impurity density in the $(a, a+da)$ range, and
$\rho((k_0,h)$ is the energy-space density of UCNs in the trap which
will be discussed below. In this case $h$ is the height $h$ of UCN over
the base of the trap  The integrations in (\ref{tau0}) are performed
over the surface of the trap and
over $k_0$ from $k_0=0$ to $k_0=k_{max}=\sqrt{2mE_{\rm max}/\hbar^2}$.
Furthermore, $4\pi\,N_1(E_{\rm max})$ is the total number of UCNs
in the trap as follows:
\begin{equation}
N_1(E_{\rm max})= \int
\rho(k_0,h)\,k_0dk_0\,dV\,,
\label{ucnnu}
\end{equation}
where the
integrations are performed over the volume of the trap, and over $k_0$
from $k_0=0$ to
$k_0=k_{max}$.
\begin{table}[h]
\begin{tabular}{|l|r|r|r|r|r|r|r|r|r|r|}\hline
$a\kappa_0$& 5.25 & 5.50 & 5.75 & 6.00 & 6.25 & 6.50 & 6.75 & 7.00 &
7.25&7.50 \\ \cline{1-11} s-wave only. & 4.819&
4.837&5.023&5.125&5.151&5.122&5.051&4.953&4.834&4.700 \\
\cline{1-11}
(s+p)-waves & 4.826&4.844&5.031&
5.136&5.165&5.140&5.077&4.992&4.907&4.940 \\ \cline{1-11}
\end{tabular}
\caption{Comparison of UCN losses/sec. $\tau_0^{-1}(E_{\rm
max},a)\times10^{5}$ at $E_{\rm max}=58$cm due to $s$ and $(s+p)$
(the last line)
interaction in the region where $p$-resonance is unavailable.}
\end{table}

A common assumption \cite{serebrov,ignat} is that
inside the trap UCNs have the isotropic velocity distribution.
Furthermore, due
to the Earth gravitation field, the UCN energy is related
with the height $h$ of UCN over the base of the
trap as $mgh=E_{\rm max}-E$, where $g$ is the acceleration of
the gravity.  So the UCN density $\rho(k_0,h)$ in both the space and
the momentum space  is given by a $\delta$-function type expression as
follows \cite{serebrov,ignat}:
\begin{equation}
\rho(k_0,h)=c_1 \delta(mgh-E_{\rm
max}+E)\,k_0\,,
\label{ddes}
\end{equation}
where $c_1$ does not depend
on the UCN wave vector and on the UCN location.
An explicit expression of
$\tau_0^{-1}(E_{\rm max},a)$ for the considered trap  is
given in Appendix B, see  eq.(\ref{lnuh}).

Fig.2 shows UCN losses in a region where resonances do not
occur. As it has been noted already,
in this case the losses are very small,
$\tau_0^{-1}(E_{\rm max},a)\sim(10^{-6}-10^{-7})$/sec.
If only a $s$-wave resonance occurs in the scattering amplitude,
then solely the $s$-wave UCN-impurity interaction is  important.
It is demonstrated by Tab.2
at $E_{\rm max}=58$ cm where the
losses due to the $s$-wave UCN-impurity interaction are compared with
the losses calculated when, in addition, the $p$-wave UCN-impurity
interaction is taken into account.

If a resonance is present in the UCN-impurity scattering amplitude,
then UCN losses increase  as it is  demonstrated by Fig.3.
When the resonance firstly occurs (it is an $s$-wave resonance)
at $E=E_{\rm max}$, a high peak arises.  Then the losses fall, but
remain rather large, $\tau^{-1}(E_{\rm max})\approx(0.5$ -- $1)
\times10^{-4}$/sec.  The losses sharply increase again when the second
resonance (it is an $p$-wave resonance) occurs.  Then the losses fall,
but remain to be $\approx(1$ -- $2)\times 10^{-4}$/sec.
In Fig.3  UCN loss probabilities
are also separately presented for off peak regions (the bottom
figures). Peak tops are not seen in Fig.3 because of the
peaks are extremely narrow.
The UCN losses within peaks are shown in Fig.4.
The UCN losses are quite large in the peak ranges, but the integral
contribution from the peak to the UCN losses is not prevailing because
of the extremely narrow width of the peak.
\begin{figure}[htbp!]
\includegraphics[width=8cm, height=6cm]{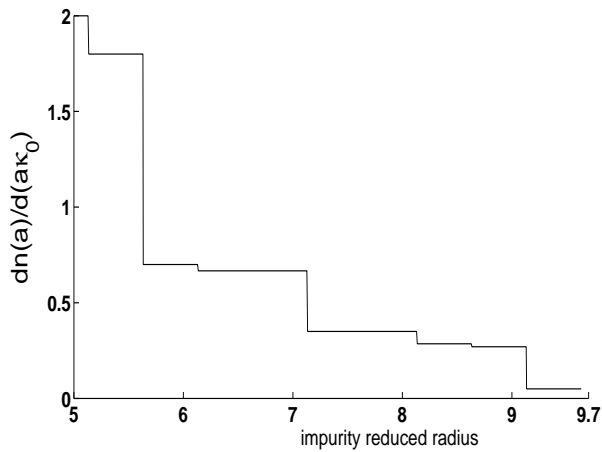}
\caption{\label{fig5}
An example of the impurity nondimensional density
$dn(a)/(da\kappa_0)$  fitting
Serebrov's data.}
\end{figure}

Experimentally measured UCN losses \cite{serebrov} can be fitted,
for instance, by means of the
impurity piecewise-smooth radius distribution
as shown in Fig.5. In this case an impurity nondimensional
density $dn(a)/(da\kappa_0)$ is given against the impurity reduced
radius $a\kappa_0$. The impurity density in the $(a, a+da)$ range
is $n_0\,\kappa_0\,da\,dn(a)/(da\kappa_0)$. As above,
$n_0=10^{14}$/cm$^3$.  Eqs. (\ref{tau0}), (\ref{tau}) and (\ref{taut0})
are employed in the calculation.  In Tab.3 UCN loss
probabilities $\tau^{-1}(E_{\rm max})$ calculated for the impurity
radius distribution in Fig.5,  are compared with the
experimental data \cite{serebrov}.
\begin{table}[h]
\begin{center}
\begin{tabular}{|c|c|c|c|c|c|c|}\hline $h_0$& 58 & 52 & 46 & 38 & 23 \\
\cline{1-6} Serebrov's data & 21.8& 17.4&15&11&8 \\ \cline{1-6}
Theor& 20.7&17.9&14.2&
11.4&7.9\\ \cline{1-6}
\end{tabular}
\end{center}
\caption{Serebrov's data (the second line) compared with  UCN losses
under an impurity radius distribution given in Fig.5 (Theor).
}
\end{table}
The impurity density $n$ within the radius range
considered is calculated as follows:
\begin{equation}
n=n_0\int_{a_{min}}^{a_{max}} \frac{dn(a)}{da}da=2.85\times10^{14}/{\rm
cm}^3\,,
\label{char}
\end{equation}
where $a_{min}=4.994/\kappa_0$ and
$a_{\rm max}=9.633/\kappa_0$. The average radius
$a_{av}$ corresponding to Fig.5, is found to be
$a_{av}=6.33/\kappa_0=575$ \AA, and
a portion $p_V$ of the volume occupied by
impurities is $p_V=0.25$.
As it was
discussed in Sec. V, a distortion of the spherical
shape impurity is potentially able  to reduces $p_V$.

\vspace{1cm}
\centerline{\Large  \bf Acknowledgments}

\vspace{0.5cm}

The author is grateful to A.P.Serebrov who attracts
his attention to
the problem of ultracold neutron losses, and for  useful
discussions. The author is grateful V.Yu. Petrov for collaboration and
useful discussions.

This work was partial supported by RSGSS-3828.2008 grant  RFFI.

\appendix
\def\thesection{Appendix \Alph{section}}
\def\theequation{\Alph{section}.\arabic{equation}}
\setcounter{equation}{0}

\section{Interference effects from impurities}
Effects from the
interference  of scattered waves are briefly discussed here. The
$s$-wave  USN-impurity interaction is only taken into account.

Impurities being $N$ in  number,
$\widetilde\psi_0({\bf r},\overrightarrow{ p},\{\alpha\})$ in
(\ref{infmat}) is
given by
\begin{equation}
\widetilde\psi_0({\bf r},\overrightarrow{ p},\{\alpha\})=
\sum_{i=1}^{i=N}A_i
\frac{e^{-\kappa|{\bf r}- {\bf r_i}|}}{|{\bf r}- {\bf r_i}|}\,,
\label{twoim}
\end{equation}
where ${\bf r_i}$ is the radius-vector of the i-th impurity center,
and the $A_i$ set
is calculated by the matching of the wave function off impurities
with the UCN wave function inside each an impurity
as follows:
\begin{equation}
e^{i\overrightarrow{ p}\cdot\overrightarrow{l}_n-\widehat\kappa(p)z_n}F_n + F_n\sum_{\neq
n}s A_s e^{-\kappa r_{ns}}/r_{ns}= A_n\,,
n=1\,\ldots,N\,;\quad
r_{nm}=|{\bf r_n-r_m}|\,,
\label{esteq}
\end{equation}
where $F_n$
is the UCN scattering amplitude on the $n$-th single
impurity.  So the $A=\{A_s\}$ column is found to be
\begin{equation}
A=
(F^{-1}-\widetilde y)^{-1}u\,,
\label{solaa}
\end{equation}
where $u=\{u_n\}$,
$F=\{F_{nm}\}$ and $\widetilde y=\{\widetilde y_{nm}\}$ are
given by
\begin{eqnarray}
u_n=e^{i\overrightarrow{p}\cdot\overrightarrow{l}_n
-\widehat\kappa(p)z_n}\,,\quad \widetilde
y_{nn}=0, \quad F_{nm}=\delta_{nm}F_m\,,
\nonumber\\
\widetilde
y_{nm}=e^{-\kappa r_{nm}}/r_{nm}\,\,\,{\rm for}\,\,n\neq m;\quad
\widetilde y_{mm}= 0\,.
\label{tilam}
\end{eqnarray}
Thus  the effect of $n$-th and of  $m$-th
impurity on each other is negligible when $F_n\widetilde y_{nm}$ and
$F_m\widetilde y_{nm}$ both are  small.  Even when $F_n=100$ km,
$F_n\widetilde y_{nm}$ is less than $0.01$ already for $|\overrightarrow{
l_n-l_m}|=1.5\times10^{-5}$ cm. The macroscopic effect is due to the
interference from those scatterers, which  separated from each other by
a macroscopic scale distances being much more than the distance above.
So $A_n$ can be approximated by the relevant amplitude of the UCN
scattering on the single isolated impurity.

To discuss the interference under the reflection
of the UCN from the trap boundary, one notes that
in the discussed case,
$\widetilde B(\overrightarrow{ q,k_0},\{\alpha\})$ is found from eq.(\ref{tif1})
with the understanding that now
the function $\widetilde\psi_0({\bf r},\overrightarrow{ p},\{\alpha\})$ is given
by (\ref{twoim}). In this case $\widetilde B(\overrightarrow{ q,k_0},\{\alpha\})$
is found to be
\begin{equation}
\widetilde B(\overrightarrow{
q,k_0},\{\alpha\})
= \sum_{s=1}^N
\sum_{s^\prime=1}^N \frac{\widehat
k_0(q_0)e^{-i\overrightarrow{ q}\cdot\overrightarrow{ l_s}} e^{-\widehat k(q)z_s} e^{i\overrightarrow{ q_0}
\cdot\overrightarrow{
l_{s^\prime}}} e^{-\widehat k(q_0)z_{s^\prime}}}{ \pi \widehat\kappa(q)(\widehat
k_0(q_0)+i\widehat\kappa(q_0))} B_{ss^\prime}\,,
\label{formbtl}
\end{equation}
where  $B_{ss^\prime}=B_{s^\prime s}$ and the  matrix $B=\{B_{ss^\prime}\}$
satisfies an equation as  follows:
\begin{equation}
B=(F^{-1}-\widehat
x)^{-1}[1-(J^{(d)}+\widehat J)B]\,.
\label{bmeq}
\end{equation}
Hence $B$ is found to be
\begin{equation}
B=(F^{-1}-\widehat x-J^{(d)}-\widehat
J)^{-1}\,,
\label{eqsb}
\end{equation}
where matrix elements of
matrices $\widehat J=\{\widehat J_{mn}\}$ and  $J^{(d)}=\{
J_{mn}^{(d)}\}$ are given by
\begin{equation}
\widehat J_{nm}= \widehat J_{mn}
=\int e^{i\overrightarrow{ q}
\cdot(\overrightarrow{ l_n}-\overrightarrow{
l_m})}\, \frac{e^{-\widehat\kappa(q)(z_n+ z_m)}\,[\widehat
k_0(q)-i\widehat\kappa(q)]}{\widehat\kappa(q)[ \widehat
k_0(q)+i\widehat\kappa(q)]}\frac{d^2q}{2\pi}\,,\quad
J^{(d)}_{sn}=-\delta_{sn}
\widetilde C_{00}^{(0)}(z_n,\kappa)\,,
\label{jssco}
\end{equation}
where $\widehat C_{00}^{(0)}(z_n,\kappa)$ is given by (\ref{coef}).
Due to a singularity at $q^2=k_0^2$ in the integrand, $\widehat J_{mn}$
decreases at $|\overrightarrow{ l_n}-\overrightarrow{ l_m}|\to\infty$, as  follows:
\begin{equation}
\widehat J_{mn}\to2(1+i)k_0e^{-\kappa_0(z_m+z_n)}
\frac{e^{k_0|\overrightarrow{
l_n}-
\overrightarrow{ l_m}|}}{ \kappa_0^2|\overrightarrow{ l_n}-
\overrightarrow{ l_m}|^2}\,,
\label{las}
\end{equation}
that is nonexponentially. To obtain (\ref{las}) one first integrates
in $\widehat J_{mn}$
over the azimuth angle $\phi$,  keeping $|\overrightarrow{ l_n}-\overrightarrow{
l_m}|\to\infty$.  The obtained integral is represented as
\begin{equation}
\widehat J_{mn}
\approx2\int_{k_0}^{\infty}e^{iq
|\overrightarrow{ l_n}-\overrightarrow{ l_m}|}\frac{e^{-\widehat\kappa(q)(z_n+
z_m)}\,[\widehat k_0(q)-i\widehat\kappa(q)]}{\widehat\kappa(q)[
\widehat k_0(q)+i\widehat
\kappa(q)]}\sqrt{\frac{q}{2\pi}}dq
+
\int_{-\infty+0}^{\infty+0}\ldots\,dq  \,,
\label{calas}
\end{equation}
where ellipses denote the integrand in the last integral. The above
integrand  is the same as in the first term. The last integral
decreases exponentially when $|\overrightarrow{ l_n}-\overrightarrow{ l_m}|\to\infty$. The
calculation of the first integral leads to (\ref{calas}).

By using (\ref{capt}), (\ref{avqu})  and
(\ref{formbtl}),
the averaged cross section $\widetilde\sigma_c(k_0,
\{\alpha\})$ of the UCN
losses in the $\widetilde y=0$ approximation  is found to be
\begin{equation}
\widetilde\sigma_c(k_0,
\{\alpha\})
=-
\frac{4\pi}{k_0^2}\sum_{m,n=1}^N
(ImJ^{(d)}+Im\widehat J)_{mn}\,|F^{-1}-J^{(d)}-\widehat
J|^{-1}_{mn}
Im F^{-1}_n\prod_{s=1}^N\frac{d^2l_s}{S}\,,
\label{avcrse}
\end{equation}
where an averaging over each $\overrightarrow{ l}_s$ is performed, as well.
Thus the $\Delta$ correction in
$\widetilde\sigma_c(k_0,\{\alpha\})$ due to the two impurity
interference  is as follows:
\begin{equation}
\Delta=-
\frac{4\pi}{k_0^2}\sum_{m,n=1}^N
\frac{ImJ^{(d)}_m\,Im F^{-1}_n|\widetilde J_{mn}|^2
+Im\widetilde
J_{mn}\,Im F^{-1}_n\,2Re[\widetilde
J_{mn}(F^{-1}_n-J^{(d)}_n)}{
\biggl|(F_m^{-1}-J^{(d)}_m)(F_n^{-1}-J^{(d)}_n)\biggl|^2}
\prod_{s=1}^N\frac{d^2l_s}{S}\,.
\label{avcrsed}
\end{equation}
From (\ref{las}), a relative correction (\ref{avcrsed}) to the
leading term of $\widetilde\sigma_c(k_0,
\{\alpha\})$ is roughly found to be
$\sim N/(S\kappa_0^2)\sim
n\,d_0/\kappa_0^2$, where $d_0$ is the length of the trap coating,
$d_0\approx 5000$ \AA, and $n$ is the impurity density. This correction
is extremely small for any reasonable $n$.

So, (\ref{avcrse}) is a sum of the losses over the UCN losses from each
a single, isolated impurity, as it is considered throughout this paper.

\section{UCN losses in a cylindrical trap}
\setcounter{equation}{0}

To obtain  $\tau_0^{-1}(E_{\rm max},a)$ in the case of interest,
cylindrical coordinates
$(\rho,\phi,z)$ are employed. In this case $z$-axis goes along the
length of the cylinder laying horizontally, $\rho$ is a minimal
distance from the given space point to $z$-axis, and $\psi$ is an angle
in the perpendicular to $z$-axis plane, $\psi=0$ at the bottommost
point of the trap.  The infinitesimal element $dS$ of the side
surface is $dS=R dzd\psi$, and for each of the butt-end,
$dS=\rho\,d\rho\,d\psi$.  The integration over $\psi$ is performed
employing $\delta$-function in (\ref{ddes}).  The integration with
respect to $z$ over the side and with respect to $\rho$ over the
butt-ends are performed without difficulties.  Then (\ref{ddes}) is
found to be
\begin{eqnarray}
\tau_0^{-1}(E_{\rm max},a)=\frac{2U}{\hbar
N(h_0)L\kappa_0}\int \frac{k_0^2}{\kappa_0^2}
\widetilde\mu_0(E,a)
\Biggl[\frac{2RL}{L_0^2\sqrt{R^2/L_0^2
-[k_0^2/\kappa_0^2-h_0/L_0+R/L_0]^2}}
\nonumber\\
+4\sqrt{
R^2/L_0^2
-[k_0^2/\kappa_0^2-h_0/L_0+R/L_0]^2}\Biggl]\frac{d
k_0^2}{\kappa_0^2}\,,
\label{taut0}
\end{eqnarray}
where
\begin{equation}
L_0=U/mg\,,\qquad h_0=E_{\rm max}/mg\,.
\label{lnuh}
\end{equation}
In the calculation of $N_1(e_{\rm max})$ one integrates over $E$
employing the $\delta$-function in (\ref{ddes}) and
integrates over $z$. In doing so $dV=\rho d\rho \psi dz$.  The
result is as it follows:
\begin{equation} N(h_0)=\int\frac{\rho
\,d\rho}{L_0^2}\,d\psi
\sqrt{(\frac{\rho}{L_0}\cos\psi-\frac{R}{L_0}\cos\psi_0)}\,,
\label{fact1}
\end{equation}
where
\begin{equation}
R-R\cos\psi_0=h_0\,.
\label{psiz}
\end{equation}
The integration in (\ref{fact1}) is performed keeping the
radicant being positive and in addition, $\psi<\psi_0$ and
$\rho<R$.  The integral can be calculated  through  Legendre
function $P_{1/2}^{-2}(1-h_0/R)$ as follows:
\begin{equation}
N(h_0)=\biggl(\frac{R}{L_0}\biggl)^{5/2} \frac{\pi}{\sqrt2}
\biggl[1-\biggl(1-\frac{h_0}{R}\biggl)^2\biggl]P_{1/2}^{-2}(1-h_0/R)\,.
\label{factex}
\end{equation}
Below eq.(\ref{factex}) is proved
for $\psi_0<0$. The calculation for $\psi_0>0$ is performed in
the same manner.

For $\psi_0<0$,  there are two integration regions in $\psi$ and $\rho$
being as follows:
\begin{eqnarray}
0<|\psi|<|\psi_0|\,,\quad 0<\rho<R \quad ({\rm
region 1})\quad{\rm and}
\nonumber\\
|\psi_0|<|\psi|<\pi\,,\quad 0<\rho|\cos\psi|<R|\cos\psi_0|\,\, (
{\rm
reg. 2}).
\label{regi}
\end{eqnarray}
The integral over each a region will
be denoted respectively
as $\widetilde N_1(h_0)/L_0^{5/2}$ and
$N_2(h_0)/L_0^{5/2}$.  The integration over $\rho$
is performed
using that the  indefinite integral
\begin{equation}
I=\int\rho \,d\rho \sqrt{(\rho\cos\psi-R\cos\psi_0)}
\label{inte}
\end{equation}
is equal to
\begin{equation}
I=
\frac{2\rho}{3\cos\psi}(\rho\cos\psi-R\cos\psi_0)^{3/2}
-
\frac{4}{15\cos^2\psi}(\rho\cos\psi-R\cos\psi_0)^{5/2} \,.
\label{intei}
\end{equation}
Hence
\begin{equation}
\widetilde N_1(h_0)=\int\,d\psi\Biggl[
\frac{2R^{5/2}}{3\cos\psi}(\cos\psi-\cos\psi_0)^{3/2}-
\frac{4R^{5/2}}{15\cos^2\psi}\biggl[(\cos\psi-\cos\psi_0)^{5/2}-
(-\cos\psi_0)^{5/2}\biggl]\Biggl]    \,,
\label{fct1}
\end{equation}
and
\begin{equation}
\widetilde N_2(h_0)=4\int\,d\psi\frac{4R^{5/2}}{15\cos^2\psi}
(-\cos\psi_0)^{5/2} \,.
\label{fct2}
\end{equation}
The second term in  (\ref{fct1}) is integrated  by
part using that $d\psi/\cos^2\psi=d\tan\psi$. Hence
\begin{equation}
\widetilde N_1(h_0)=\frac{2R^{5/2}}{3}\int\,d\psi\,\cos\psi
(\cos\psi-\cos\psi_0)^{3/2}
- \frac{8R^{5/2}}{15}\sin\psi_0
(-\cos\psi_0)^{3/2}\,.
\label{ffct1}
\end{equation}
Furthermore,
\begin{equation}
\frac{2R^{5/2}}{3}\int\,d\psi\,\cos\psi
(\cos\psi-\cos\psi_0)^{3/2}
=
\pi\frac{R^{5/2}}{\sqrt2}\sin^2\psi_0
P^{-2}_{1/2}(\cos\psi_0)\,,
\label{ffcct1}
\end{equation}
where $P^{-2}_{1/2}$ is the
Legendre function. The $\widetilde N_2(h_0)$ integral is easy calculated,
and as a final result, (\ref{factex}) arises.

\end{document}